\documentclass[a4paper,11pt]{article}
\pdfoutput=1 

\usepackage{jheppub} 

\usepackage{subcaption}

\usepackage{diagbox}

\usepackage{multirow}
\usepackage[table]{xcolor}
\usepackage{mathtools}

\newcommand{\lran}[1]{{\langle {#1}\rangle}}
\newcommand{\bfk}{{{\bf k}}}

\newcommand{\dep}{\delta \phi}
\newcommand{\bp}{\bar{\phi}}

\title{
Heavy Spinning Particles from Signs of Primordial Non-Gaussianities: Beyond the Positivity Bounds} 

\author[a]{Suro Kim,}

\author[a]{Toshifumi Noumi,}

\author[a]{Keito Takeuchi}

\author[b]{and Siyi Zhou}

\affiliation[a]{Department of Physics, Kobe University, Kobe 657-8501, Japan}
\affiliation[b]{Department of Physics and Jockey Club Institute for Advanced Study,
 Hong Kong University of Science and Technology, Hong Kong}

\emailAdd{s-kim@stu.kobe-u.ac.jp}

\emailAdd{tnoumi@phys.sci.kobe-u.ac.jp}

\emailAdd{181s113s@stu.kobe-u.ac.jp}

\emailAdd{szhouah@ust.hk}

\preprint{KOBE-COSMO-19-09}

\abstract{
Within the so-called cosmological collider program, imprints of new particles on primordial non-Gaussianities have been studied intensively. In particular, their non-analytic features in the soft limit provide a smoking gun for new particles at the inflation scale. While this approach is very powerful to probe particles of the mass near the Hubble scale, the signal is exponentially suppressed for heavy particles. In this paper, to enlarge the scope of the cosmological collider, we explore a new approach to probing spins of heavy particles from signs of Wilson coefficients of the inflaton effective action and the corresponding primordial non-Gaussianities. As a first step, we focus on the regime where the de Sitter conformal symmetry is weakly broken. It is well known that the leading order effective operator $(\partial_\mu\phi\partial^\mu\phi)^2$ is universally positive as a consequence of unitarity. In contrast, we find that the sign of the six derivative operator $(\nabla_\mu\partial_\nu\phi)^2(\partial_\rho\phi)^2$ is positive for intermediate heavy scalars, whereas it is negative for intermediate heavy spinning states.
Therefore, under the assumption of tree-level UV completion, the sign can be used to probe spins of heavy particles generating the effective interaction.
We also study phenomenology of primordial non-Gaussianities thereof.}

\begin{document} 
\setcounter{tocdepth}{2}
\maketitle
\flushbottom

\section{Introduction}
\setcounter{equation}{0}

The energy scale of inflation could be as high as $10^{13}$ GeV, hence it would be a phenomenon at the highest energy scale we may explore. In particular, primordial non-Gaussianities can be thought of as an extremely high energy particle collider. For example, their non-analytic behaviors in the soft limit are associated with on-shell particle creations and thus provide a direct evidence of new particles at the inflation scale~\cite{Chen:2009zp,Baumann:2011nk,Noumi:2012vr,Arkani-Hamed:2015bza}. Just like resonance signals at particle colliders, it is very powerful to probe particles with the mass $m$ near the Hubble scale $H$. Such an approach is dubbed the cosmological collider program and has been studied intensively~\cite{Chen:2009we,Baumann:2011nk,Noumi:2012vr,Chen:2009zp,Assassi:2012zq,Sefusatti:2012ye,Norena:2012yi,Emami:2013lma,Liu:2015tza,Arkani-Hamed:2015bza,Dimastrogiovanni:2015pla,Schmidt:2015xka,Chen:2015lza,Delacretaz:2015edn,Bonga:2015urq,Flauger:2016idt,Lee:2016vti,Delacretaz:2016nhw,Meerburg:2016zdz,Chen:2016uwp,Chen:2016hrz,Kehagias:2017cym,An:2017hlx,Tong:2017iat,Iyer:2017qzw,An:2017rwo,Kumar:2017ecc,Riquelme:2017bxt,Franciolini:2017ktv,Saito:2018omt,Cabass:2018roz,Wang:2018tbf,Dimastrogiovanni:2018uqy,Bordin:2018pca,Chua:2018dqh,Arkani-Hamed:2018kmz,Kumar:2018jxz,Goon:2018fyu,Wu:2018lmx,Anninos:2019nib,McAneny:2019epy,Li:2019ves}. For heavy particles, on the other hand, the signal is exponentially suppressed by the Boltzmann factor. For scalars, this factor reads $e^{-\pi\sqrt{\frac{m^2}{H^2}-\frac{9}{4}}}$, which is as small as $10^{-4}$ even for $m=3H$! Therefore, it is desirable to develop a new approach to probing heavy particles to enlarge the scope of the cosmological collider\footnote{
In contrast to ordinary colliders, we cannot build a new cosmological collider with a higher energy $H$!}.

\medskip
Now let us recall our history of particle physics, where not only resonance signals, but also detailed studies of low-energy effective interactions have been useful to probe new particles. A typical example is the prediction of weak bosons, where angular dependence of the Fermi interactions played an important role. Following the history, we would like to apply a similar idea to inflaton effective interactions and primordial non-Gaussianities. Recall that the inflaton enjoys a shift symmetry under the slow-roll approximation. Its effective Lagrangian then reads \cite{Weinberg:2008hq}
\begin{align}
\mathcal{L}_\phi=-\frac{1}{2}(\partial_\mu\phi)^2+\frac{\alpha}{\Lambda^4} (\partial_\mu\phi\partial^\mu\phi)^2+\ldots\,,
\end{align}
where the dots stand for higher dimensional operators and the cutoff $\Lambda$ is typically associated to the mass of intermediate heavy states. Remarkably, it is well known that $\alpha$ is always positive in a wide class of theories as a consequence of unitarity and analyticity of scattering amplitudes~\cite{Adams:2006sv} (see, e.g., \cite{Baumann:2015nta,Cheung:2016yqr,deRham:2017avq,deRham:2017imi,deRham:2017zjm,Bellazzini:2017fep,deRham:2017xox,deRham:2018qqo,Hamada:2018dde,Chen:2019qvr,Bellazzini:2019xts,Herrero-Valea:2019hde,Baumann:2019ghk} for recent applications). 
While the positivity is an elegant consistency condition on IR effective theories\footnote{
In other words, if experiments find violation of the positivity bounds, we have to change our approach to UV completion in a drastic way, which is also an interesting possibility.
}, detailed informations of the UV theory, e.g., spins of heavy states, are obscured at the cost of universality.

\medskip
In this paper we would like to go beyond the positivity and develop an approach to probing spins of heavy states from signs of inflaton effective interactions and the corresponding primordial non-Gaussianities. As a first step, we focus on the regime where the de Sitter conformal symmetry is weakly broken\footnote{
The nonlinearity parameter $f_{NL}$ for three-point functions is generically smaller than $\mathcal{O}(1)$ in this regime. At this cost, we may utilize developments on Lorentz invariant four-point amplitudes instead. The same remark applies to the elegant works~\cite{Arkani-Hamed:2015bza,Arkani-Hamed:2018kmz} based on the de Sitter invariant four-point functions. In order to study the regime $f_{NL}>\mathcal{O}(1)$, we need to incorporate more general higher derivative operators such as $(\partial_\mu\phi\partial^\mu\phi)^n$ ($n\geq3$) or work in the effective field theory of primordial perturbations~\cite{Cheung:2007st}. }. In this regime, we may focus on operators with four inflaton fields for the study of primordial three-point and four-point functions:
\begin{align}
\label{EFT_phi}
\mathcal{L}_\phi=-\frac{1}{2}(\partial_\mu\phi)^2+\frac{\alpha}{\Lambda^4} (\partial_\mu\phi\partial^\mu\phi)^2+\frac{\beta}{\Lambda^6}(\nabla_\mu\partial_\nu\phi)^2(\partial_\rho\phi)^2+\ldots\,,
\end{align}
which provides the most general effective Lagrangian up to six derivatives after using tree-level equations of motion to remove operator degeneracy. In contrast to $\alpha$, the sign of $\beta$ cannot be fixed by analyticity and unitarity essentially because the corresponding four-point amplitude vanishes in the forward limit. In other words we may have a chance to probe spins of heavy states from its sign. By generalizing the analysis in~\cite{Adams:2006sv} to non-forward amplitudes, we show that
\begin{enumerate}
\item $\beta>0$ for intermediate heavy scalars, whereas $\beta<0$ for heavy spinning states with spin $l=2,4,6,...$, as long as four-point amplitudes are bounded as $<s^2$ in the high energy limit ($s$ is the standard Mandelstam variable).
\end{enumerate}
Note that our conclusion is about intermediate ``states," which can be multi-particle states in general. See Appendix~\ref{loopdiagramappendix} for more comments.
While the argument is applicable to a wide class of theories, it cannot be applied directly to the effective coupling generated by the exchange of KK gravitons, which is one of the main targets in the cosmological collider program\footnote{See, e.g.,~\cite{Kumar:2018jxz} for non-analytic behaviors of primordial non-Gaussianities generated by KK gravitons. While the signature discussed there provides a direct evidence of massive spin $2$ particles, it is exponentially suppressed by the Boltzmann factor unfortunately. On the other hand, the heavy mass suppression of the signal we discuss in this paper is polynomial $\sim(H/m_{\rm KK})^4$.}, essentially because it involves gravitational dynamics (see Sec.~\ref{kk} for details).
Interestingly, however, we find that 
\begin{enumerate}
\setcounter{enumi}{1} 
\item the six derivative operator generated by the KK graviton exchange has a negative coefficient $\beta<0$.
\end{enumerate}
This shows that a non-positive coefficient $\beta\leq0$ is a sign of heavy spinning particles with the spin $l=2,4,6...$ in a wide class of theories in our interests. This is one of the main results in our paper. See Table~\ref{table:summary} for a summary. We also study phenomenology of primordial non-Gaussianities thereof.

\begin{table}[] \centering
	\begin{tabular}{|l|l|l|l|ll}  
		\cline{1-4}
		&Four derivative&Six derivative& Dominant UV States &    \\ \cline{1-4}
		\multirow{3}{*}{IR} & \multirow{3}{*}{$a_{2,0}>0$,\quad $\alpha>0$}&$a_{2,1}<0$,\quad $\beta>0$ & Scalar &    \\ \cline{3-4}
		& &$a_{2,1}=0$,\quad $\beta=0$ & Exact Cancellation &    \\ \cline{3-4}
		& &$a_{2,1}>0$,\quad $\beta<0$ & Spin $l=2,4,6\ldots$ &    \\ \cline{1-4} 
	\end{tabular}
\caption{The sign of $\alpha$ in the low energy EFT (or equivalently the coefficient  $a_{2,0}$ of $s^2$ in the low-energy four-point scattering) is universally positive. On the other hand, the sign of $\beta$ (or equivalently the coefficient  $a_{2,1}$ of $s^2t$ in the low-energy four-point scattering) reflects which UV states provide dominant contributions to low-energy scattering. The conclusion summarized in the table applies to any weakly coupled, Lorentz-invariant, analytic and unitary UV completion as well as effective interactions mediated by KK gravitons. }
	\label{table:summary}
\end{table} 

\medskip
The organization of the paper is as follows. 
In Sec.~\ref{sign} we first study the relation between the sign of the six derivative operator $(\nabla_\mu\partial_\nu\phi)^2(\partial_\rho\phi)^2$ and the spins of intermediate heavy states. We then study  shapes of four-point functions (Sec.~\ref{4pt}) and three-point functions (Sec.~\ref{3pt}) to clarify if those generated by the six derivative operator are distinguishable from the four derivative ones. We  conclude in  Sec.~\ref{conclusion} with a discussion of future directions.

\section{Spin-dependence of Six Derivative Operators }
\label{sign}
In this section we clarify the spin-dependence of the six derivative operator. In Sec.~\ref{uv} we first introduce a general basis of four-point amplitudes consistent with the Froissart-Martin bound neglecting gravitational effects. We then utilize it in Sec.~\ref{beyond} to show that the sign of the six derivative operator is positive (negative) for intermediate heavy scalars (spinning states). In Sec.~\ref{string} we consider open superstring theory as an example for UV completion. There we demonstrate that the six derivative operator vanishes as a consequence of cancellation between intermediate scalars and spinning particles. Besides, in Sec.~\ref{kk}, we study amplitudes mediated by the Kaluza-Klein graviton, to which we cannot apply the results in Secs.~\ref{uv}-\ref{beyond} in general because of gravitational effects, but it is of great interests in the cosmological collider program. A remark on loops is also given in Appendix~\ref{loopdiagramappendix}.

\subsection{A basis of general UV amplitudes }\label{uv}

\begin{figure}[t] 
	\centering 
	\includegraphics[width=7cm]{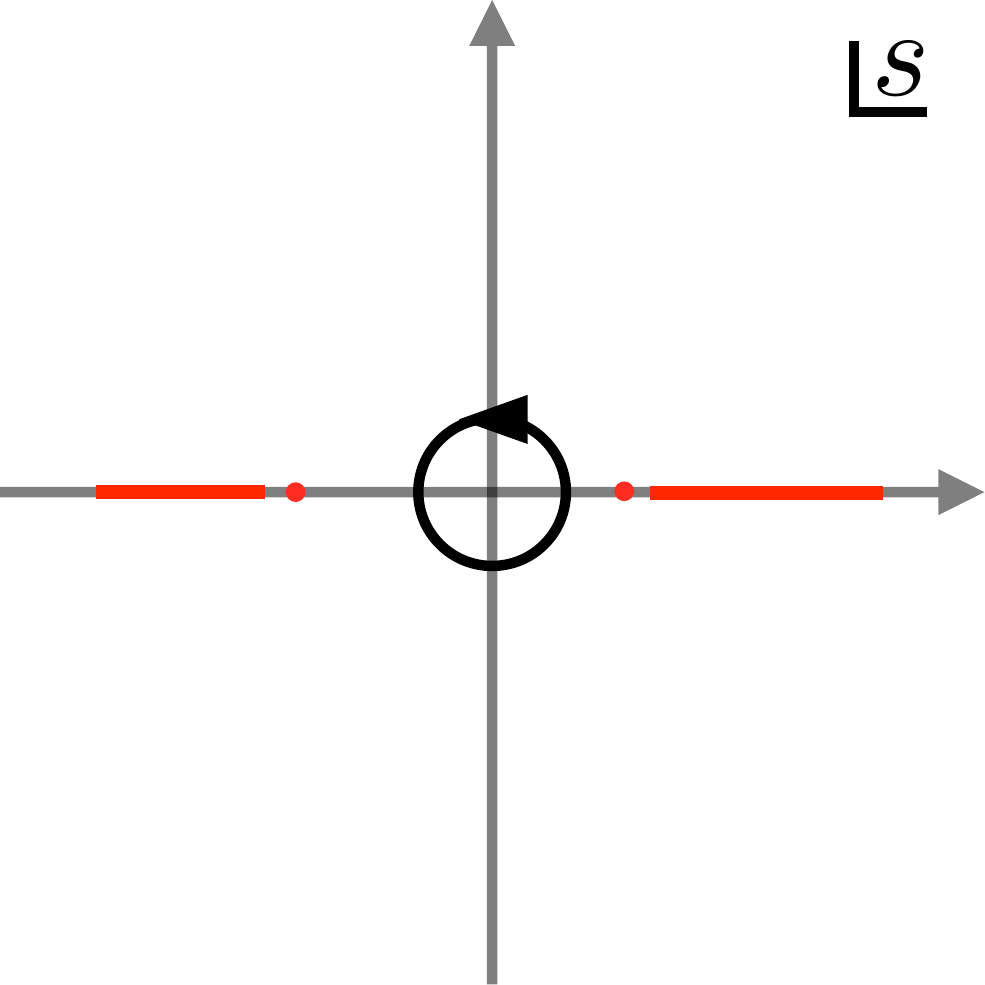}   \quad\quad
	\includegraphics[width=7cm]{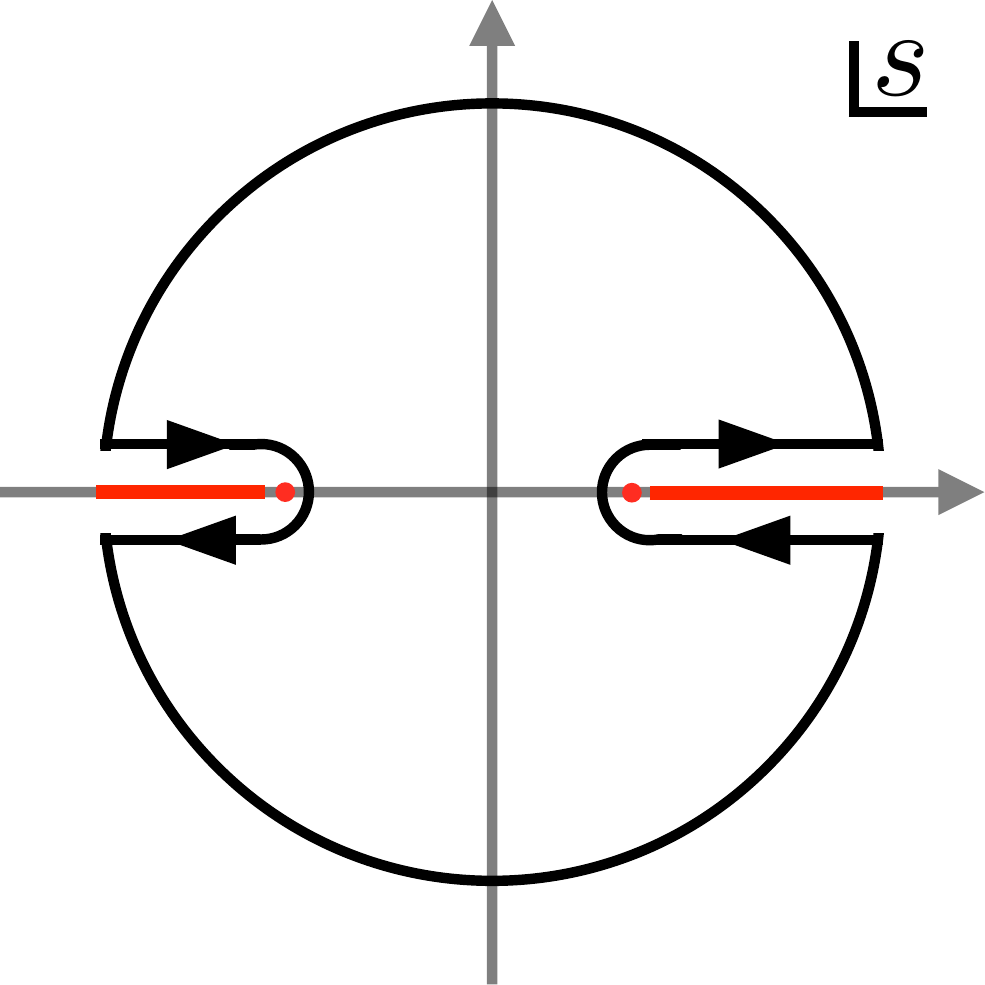}
	\caption{Integration contour and analytic structure:
	The red lines represent branch cuts associated to multi-particle states generated by loops, whereas the red dots are single poles associated to single-particle states. The left panel is the integration contour used to read off the IR coefficients. By deforming it into the one on the right panel, we may clarify how the UV data is encoded into the IR coefficients.} \label{contourintegral}
\end{figure}

Let us consider four-point scattering amplitudes $M(s,t)$ of identical massless scalars and elaborate on the relation between IR coefficients and the UV spectrum. For this purpose, we generalize the analysis in~\cite{Adams:2006sv} to non-forward amplitudes (see also~\cite{deRham:2017avq}).
First, from the IR point of view, it is convenient to expand the amplitude $M(s,t)$ in Mandelstam variables $s$ and $t$ as
\begin{align}\label{SA0residue}
M(s,t)=\sum_{p,q=0}^\infty a_{p,q} s^p t^q\,,
\end{align}
where we assumed that gravity is subdominant to neglect massless cuts and the massless graviton pole. The coefficient function of $s^p$ can then be evaluated by the contour integral,
\begin{align}\label{contourintegralmaster}
\sum_{q=0}^\infty a_{p,q} t^q=\oint\frac{ds}{2\pi i}\frac{M(s,t)}{s^{p+1}}\,,
\end{align}
where the integration contour is defined such that it encircles the origin $s=0$ and the integrand is analytic inside the contour except for the origin (see the left panel of Fig.~\ref{contourintegral}). Also we take $t$ infinitesimal to avoid unphysical poles.

\medskip
Next, we deform the contour as depicted in the right panel of Fig.~\ref{contourintegral} by assuming that the amplitude is analytic away from the real axis on the physical sheet. If the amplitude is bounded as $|M(s,t)|<|s|^p$ at UV, the integral at infinity vanishes to obtain
\begin{align}\label{contourdeformed}
\oint\frac{ds}{2\pi i}\frac{M(s,t)}{s^{p+1}}=
\left(\int_{m_0^2}^\infty+\int_{-\infty}^{-m_0^2-t}\right)\frac{ds}{2\pi i}\frac{M(s+i\epsilon,t)-M(s-i\epsilon,t)}{s^{p+1}}\,,
\end{align}
where $\epsilon$ is an infinitesimal positive constant and $m_0$ is the mass of the lightest intermediate (massive) state. 
In general the amplitude contains single poles associated with tree-level exchange and  branch cuts associated with loops, so that the integral is given by
\begin{align}\label{masterfomula}
\nonumber
\text{r.h.s. of \eqref{contourdeformed}}=&\sum_{n} \left[ \frac{g_{n}^2 P_{l_n}\big(1+\frac{2t}{m_n^2}\big)}{(m_n^2)^{p+1}} - \frac{g_{n}^2 P_{l_n}\big(1+\frac{2t}{m_n^2}\big)}{(-m_n^2-t)^{p+1}} \right]
\\
&+\sum_l\int_{s_*}^\infty dM^2\rho(M^2,l)\left[ \frac{P_{l}\big(1+\frac{2t}{M^2}\big)}{(M^2)^{p+1}} - \frac{P_{l}\big(1+\frac{2t}{M^2}\big)}{(-M^2-t)^{p+1}} \right]\,,
\end{align}
where the first line and the second line correspond to single poles and branch cuts, respectively. $P_n(z)$ is the Legendre polynomial, $m_n$ and $l_n$ are the mass and spin of the intermediate particle labeled by $n$, and $g_n$ characterizes the three-point amplitude of two massless scalars and one massive particle $n$.
$\rho(M^2,l)$ is a spectral density for the transition amplitude from two massless particles to multi-particle states with the energy $M$ and total angular momentum $l$. 
$s_*$ is the minimum energy squared of intermediate multi-particle states.
For notational simplicity, we write Eq.~\eqref{masterfomula} as
\begin{align}
\text{r.h.s. of \eqref{contourdeformed}}=&\sum_{n} \left[ \frac{g_{n}^2 P_{l_n}\big(1+\frac{2t}{m_n^2}\big)}{(m_n^2)^{p+1}} - \frac{g_{n}^2 P_{l_n}\big(1+\frac{2t}{m_n^2}\big)}{(-m_n^2-t)^{p+1}} \right] \,,
\end{align}
where one should understand that $\sum_n$ stands for both of the discrete sum for tree-level exchange and the integration (continuous sum) for loops.
Note that since we are considering identical external massless scalars, the intermediate state can have an even spin only $l_n=0,2,4,...$ as required by the exchange symmetry (for simplicity, we use ``spins" to denote total angular momenta of multi-particle states as well). See, e.g.,~\cite{Arkani-Hamed:2017jhn} for details.

\medskip
We now conclude that the IR coefficients are given in terms of the UV data as
\begin{align}
\label{apq_Legendre}
\sum_{q=0}^\infty a_{p,q} t^q=\sum_{n} \left[ \frac{g_{n}^2 P_{l_n}\big(1+\frac{2t}{m_n^2}\big)}{(m_n^2)^{p+1}} - \frac{g_{n}^2 P_{l_n}\big(1+\frac{2t}{m_n^2}\big)}{(-m_n^2-t)^{p+1}} \right]\,.
\end{align}
Note that this relation is applicable only for $p$ satisfying $|M(s,t)|<|s|^p$ at UV. On general grounds, we assume that the UV amplitude is bounded as $|M(s,t)|<|s|^2$, which is required by the Froissart-Martin bound \cite{Froissart:1961ux} for example\footnote{
To be precise, the Froissart-Martin bound is applicable only in gapped theories and in the regime $0\leq t<4m^2$ ($m$ is the mass of external particles). However, its use can be justified by turning on a tiny mass of the scalar as a regulator, which does not change the conclusion of our analysis. Also note that the inflaton indeed has a tiny mass which is suppressed due to the slow-roll conditions, even though the mass is negligible as long as we are interested in non-Gaussianities generated by higher derivative operators.}. Hence, the relation is applicable only for $p\geq 2$. It is also instructive to reorganize the amplitude as
\begin{align}\label{scatteringamplitudemaster}
M(s,t)=\sum_{n} g_{n}^2\,P_{\ell_n}\!\!\left(1+\frac{2t}{m_n^2}\right)
\left[
\frac{1}{m_n^2-s}+\frac{1}{m_n^2+s+t}
\right]
+\alpha_0(t)+\alpha_1(t)s\,.
\end{align}
Here we introduced $\displaystyle\alpha_p(t)=\sum_{q=0}^\infty a_{pq}t^q$ for $p=0,1$, which cannot be determined by the contour deformation argument in the above. Eq.~\eqref{scatteringamplitudemaster} provides a basis of general UV amplitudes which reproduce the correct factorization and satisfy the Froissart-Martin bound.

\subsection{Positivity and the beyond }
\label{beyond}

We then explore imprints of the UV data on the IR coefficients $a_{p,q}$. First, from Eq.~\eqref{apq_Legendre}, the IR coefficients read
\begin{align}
a_{p,q}=\sum_n\frac{g_n^2}{(m_n^2)^{p+q+1}}\left[ \frac{(l_n-q+1)_{2q} }{(q!)^2  }
+
\sum_{k=0}^q(-1)^{p+q+k} \frac{(l_n-k+1)_{2k} (p+1)_{q-k}}{(k!)^2(q-k)!  }
\right]\,,
\end{align}
where $(x)_n=\Gamma(x+n)/\Gamma(x)=x(x+1)\cdots (x+n-1)$ is the shifted factorial (also called the Pochhammer symbol) and we used
\begin{align}
P_n(z)=\sum_{k=0}^n\frac{(-n)_k(n+1)_k}{(k!)^2}\left(\frac{1-z}{2}\right)^k
=\sum_{k=0}^n\frac{(n-k+1)_{2k}}{(k!)^2}\left(\frac{z-1}{2}\right)^k\,.
\end{align}
Let us next take a closer look at the first few orders in the $t$ expansion:
\begin{enumerate}
\setcounter{enumi}{-1}
\item $\mathcal{O}(t^0)$ ($q=0$)

First, the leading order coefficients in the $t$ expansion are given by
\begin{align}
\label{a_p,0}
a_{p,0}=
\left\{\begin{array}{cl}\displaystyle 2\sum_n\frac{g_n^2}{(m_n^2)^{p+1}} & \quad\text{for even $p$}\,, \\0 &\quad\text{for odd $p$}\,,\end{array}\right.
\end{align}
which is nothing but the well-known positivity bounds on the $\mathcal{O}(s^{2n})$ coefficients of forward amplitudes~\cite{Adams:2006sv}. This bound is universal and elegant, but detailed information such as spins of the UV states is obscured at the cost of universality.

\item $\mathcal{O}(t^1)$ ($q=1$)

Similarly, the next-to-leading order coefficients read
\begin{align}
\label{a_p,1}
a_{p,1}=
\left\{\begin{array}{cl}\displaystyle \sum_n\frac{g_n^2}{(m_n^2)^{p+2}} \left(2l_n^2+2l_n-p-1\right)& \quad\text{for even $p$}\,, \\ \displaystyle(p+1) \sum_n\frac{g_n^2}{(m_n^2)^{p+2}} &\quad\text{for odd $p$}\,.\end{array}\right.
\end{align}
The point here is that the sign of $a_{p,1}$ for even $p$ depends on the spin $l_n$ of the intermediate states. More explicitly, we find that intermediate states with the spin higher (lower) than the critical value $l_*=(\sqrt{2p+3}-1)/2$ gives a positive (negative) contribution to $a_{p,1}$. In particular, the critical value is $l_*\sim0.8$ for $p=2$. In terms of the effective Lagrangian~\eqref{EFT_phi}, it means that $\beta$ is positive only for intermediate scalars as we discuss shortly. This six derivative operator is the first nontrivial operator in the derivative expansion which may be used to probe spins of intermediate states. On the other hand, $a_{p,1}$ for odd $p$ is universally positive. 

\item $\mathcal{O}(t^2)$ ($q=2$)

As a next example, let us consider the $\mathcal{O}(t^2)$ term, for which we find
\begin{align}
a_{p,2}=
\left\{\begin{array}{cl}\displaystyle \frac{1}{2}\sum_n\frac{g_n^2}{(m_n^2)^{p+3}} \left(l_n^2-p-2\right)\left(l_n^2+2l_n-p-1\right)& \quad\text{for even $p$}\,, \\ \displaystyle\frac{1}{2}(p+1) \sum_n\frac{g_n^2}{(m_n^2)^{p+3}}\left(2l_n^2+2l_n-p-2\right) &\quad\text{for odd $p$}\,.\end{array}\right.
\end{align}
For odd $p$, there exists a critical spin $l_*=(\sqrt{2p+5}-1)/2$ below (above) which $a_{p,q}$ is negative (positive). On the other hand, for even $p$, there exists a window $\sqrt{p+2}-1<l_n<\sqrt{p+2}$ inside (outside) which $a_{p,q}$ is negative (positive). Note that it has a width $1$, so that at most one value of spin can be inside the window.  For example, the window is $1<l_n<2$ for $p=2$, hence $a_{2,2}=0$ for spin 2, whereas $a_{2,2}>0$ for other even spins. Similarly, spin $2$ is inside the window for $p=4,6$.

\item Higher orders

It is straightforward to go higher in a similar way. First, $a_{p,q}$ is a polynomial in $l_n$ of order $2q$ for even $p$ and order $2q-2$ for odd $p$. It is positive when the spin $l_n$ is sufficiently large for fixed $p,q$. On the other hand, the contribution  from a scalar $l_n=0$ is simply $(-1)^{p+q}(p+1)_q/q!(m_n^2)^{p+q+1},$\footnote{Note that this expression is applicable only for $q\neq0$.} hence the sign depends only on $(-1)^{p+q}$. The sign for general spin depends on details of the polynomial in $l_n$ and the phase structure will be richer for higher $q$. However, we will not go into more details because such higher order terms are not easy to probe phenomenologically, leaving it for future work.

\end{enumerate}
To summarize, we have elaborated on the relation between the sign of IR coefficients $a_{p,q}$ and the spin of intermediate heavy states. The result can also be rephrased in terms of EFT coefficients as follows: For example, four-point scattering amplitudes can be evaluated from the effective Lagrangian~\eqref{EFT_phi} as
	\begin{align}
	M(s,t) = \frac{4\alpha}{\Lambda^4} (s^2+t^2+st) - \frac{3\beta}{\Lambda^6}(s^2 t+s t^2)+\ldots\,,
	\end{align}	
	where the dots stand for higher derivative terms negligible at low-energy. We then find
	\begin{align}
	a_{2,0} = \frac{4\alpha}{\Lambda^4}, \quad a_{2,1} = -3 \frac{\beta}{\Lambda^6}~.
	\end{align}
Therefore, our analysis on $a_{2,1}$ implies that the coefficient $\beta$ of the six derivative operator $(\nabla_\mu\partial_\nu\phi)^2(\partial_\rho\phi)^2$ is positive only for intermediate scalars. In other words, detection of negative or vanishing $\beta$ is a smoking gun of spinning heavy states with the even spin $l_n=2,4,...$\footnote{
To unitarize theories with a massive particle of spin 2 or higher in a weakly coupled regime, one would expect an infinite Regge tower of higher spin particles.
If it is a universal requirement of weakly coupled UV completion, we could think of a non-positive $\beta$ as a sign of the infinite higher spin tower.}. This is one of our main results.

\medskip
It should be noted again that our results are about the spins (or total angular momenta) of heavy states, which can be multi-particle states when loop effects are dominant. However, in the case of inflation, the inflaton has to enjoy the approximate shift symmetry. As a result, the loop effects are generically suppressed by either weak couplings or slow-roll parameters as we discuss in Appendix \ref{App_loop}. In such a case, we can use the sign of effective interactions to probe the spins of the intermediate heavy {\it particles} exchanged at the tree-level without swamped by multi-particle states associated with loops.

\subsection{Open superstring amplitudes }\label{string}

In the previous subsection we have shown that the non-positivity of the coefficient $\beta$ of the six derivative operator $(\nabla_\mu\partial_\nu\phi)^2(\partial_\rho\phi)^2$  is a signature of higher spin states. However, it does not mean that $\beta\leq0$ holds in all theories with higher spins because intermediate scalars may dominate to make $\beta$ positive. Therefore, it is useful to demonstrate that there indeed exists a UV completion with higher spins and a non-positive $\beta$. In this subsection we study open superstring theory to provide a concrete example for such a UV completion.

\medskip
Let us consider four-point scattering of identical massless scalars  (an extra-dimensional component of the gauge boson) localized on a D3 brane in the open superstring. The corresponding disk amplitude reads\footnote{
We suppressed an overall positive coefficient associated with normalization of the string coupling.} (see, e.g.,~\cite{Schlotterer:2011psa}) 
\begin{align}
\label{M_string}
M(s,t)  =\left(s^2+t^2+u^2\right) \left[
\frac{B(-s,-t)}{s+t}
+\frac{B(-t,-u)}{t+u}
+\frac{B(-u,-s)}{u+s}
\right]
\,,
\end{align}
where $u=-s-t$ and $B(a,b)$ is the Euler beta function,
\begin{align}
B(a,b)=\int_0^1 dx x^{a-1} (1-x)^{b-1}
=\frac{\Gamma(a)\Gamma(b)}{\Gamma(a+b)}\,.
\end{align}
We also took the unit in which the open string spectrum is given by $m^2=0,1,2,...$. At IR, we may expand the amplitude in the Mandelstam variables as
\begin{align}
M(s,t)  =\pi^2\left(s^2+st+t^2\right)
+\frac{\pi^4}{12}\left(s^2+st+t^2\right)^2
+\ldots
\,,
\end{align}
where the dots stand for fifth and higher orders in Mandelstam variables. We find that the six derivative operator vanishes, which means that there exists an exact cancellation between scalars and higher spins.

\begin{figure}[htbp] 
	\centering 
	\includegraphics[width=15cm]{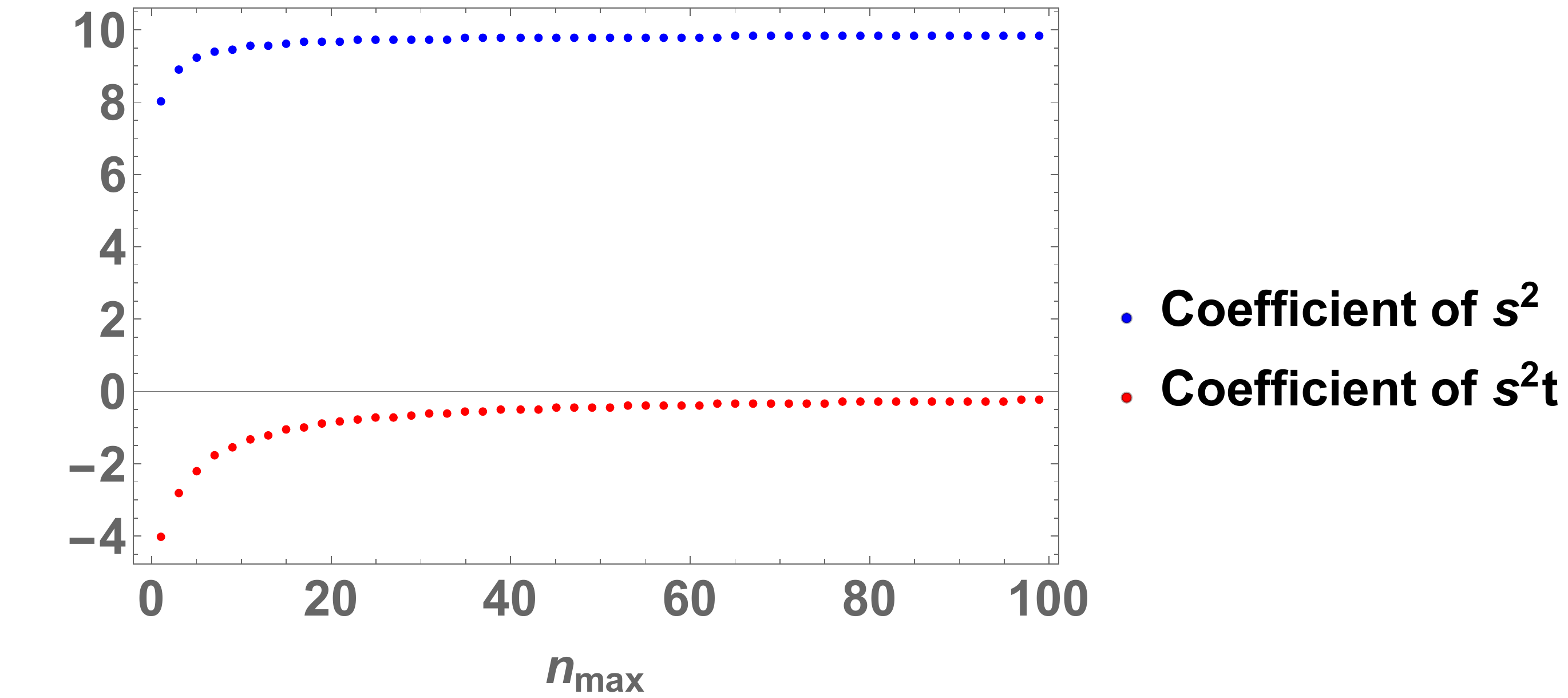}  
	\caption{The coefficients of $s^2$ (blue) and $s^2 t$ (red) obtained after summing up over $n$ from $1$ to $n_{\rm max}$:
	As the positivity bound implies,  each sector labeled by $n$ contributes to the $s^2$ coefficient positively. On the other hand, the $n=1$ sector gives a negative contribution to the $s^2t$ coefficient, which is canceled by positive contributions from $n\geq3$ to obtain a vanishing coefficient in the limit $n_{\rm max}\to\infty$ representing the open superstring amplitude. } \label{sumover}
\end{figure}

\medskip
It is also instructive to explicitly see how the cancellation happens between scalars and higher spins by comparing the open string amplitude~\eqref{M_string} with the master formula \eqref{scatteringamplitudemaster} to read off the cubic couplings $g_n$. First, the amplitude~\eqref{M_string} has single poles associated with particles of the mass squared $m^2=1,3,5,...$. For example, the residue of the $s$-channel pole reads 
\begin{align}
	{\rm Res}_{s\to n} M(s,t) & = -\frac{4 \left(t^2+nt+n^2\right) (t+1)_{n-1}}{n!}
	\quad
	(n=1,3,5,...) \,.
\end{align}
It is a polynomial in $t$ of order $n+1$, which consists of contributions from intermediate particles with spin up to $n+1$. More explicitly, we expand it by Legendre polynomials as
\begin{align}
{\rm Res}_{s\to n} M(s,t) & =-\sum_{l} g_{nl}^2\,P_{l}\!\left(1+\frac{2t}{n}\right) \,,
\end{align}
where $g_{nl}$ characterizes cubic coupling of two massless scalars and the massive particle of the mass squared $m^2=n$ and spin $l$\footnote{To be precise, the string spectrum generically contains multiple species of particles for given $n$ and $l$, hence $g_{nl}^2$ is a summation over the cubic coupling squared of them.}. For example, the $n=1$ sector is given by
\begin{align}
g_{10}^2=\frac{10}{3}\,,
\quad
g_{12}^2=\frac{2}{3}\,.
\end{align}
Note that odd spins do not appear as an intermediate particle because of the exchange symmetry of identical scalars mentioned earlier. We then find that the $n=1$ sector gives a negative contribution to the coefficient $a_{2,1}$ of $s^2t$:  
\begin{align}
a_{2,1}
=\sum_{n,l}\frac{g_{nl}^2}{n^4}\left(2l^2+2l-3\right)
\ni \sum_{l=0,2}g_{1l}^2\left(2l^2+2l-3\right)=-4\,,
\end{align}
where we used Eq.~\eqref{a_p,1}. In other words, the scalar contribution dominates over the higher spin one in this sector. It is straightforward to generalize the argument to general $n$. $g_{nl}^2$ for the first several orders are given in Table~\ref{table1}. We find that for $n=3,5,...$ higher spins dominate over scalars to obtain a positive contribution to $a_{2,1}$. In Fig.~\ref{sumover}, for illustration, we provide a plot for the coefficient of $s^2t$, 
\begin{align}
\sum_{n=1}^{n_{\rm max}}\sum_{l=0}^{n+1}\frac{g_{nl}^2}{n^4}\left(2l^2+2l-3\right)\,,
\end{align}
obtained after summing up over $n$ from $1$ to $n_{\rm max}$, which shows that the coefficient approaches to $0$ asymptotically as we increase $n_{\rm max}$.

\begin{table}\centering 
\begin{tabular}{|c|c|c|c|c|c|c|}
	\hline
	\diagbox[dir=NW]{$l$}{$g_{nl}^2$}{$n$} 
	&  1 &  3  &  5  &  7  &  9 \\	\hline	
0	&  $ \frac{10}{3}$ &$ \frac{14}{5}$ & $\frac{425}{168} $& $\frac{76517}{32400}$ & $\frac{552711}{246400}$ \\	\hline	  
2	& $\frac{2}{3}$ & $\frac{59}{7}$ & $\frac{4805}{504}$ & $\frac{693133}{71280}$ & $\frac{789591}{81536}$ \\\hline	
4	&0 & $\frac{27}{35}$ & $\frac{13625}{1848}$ & $\frac{5363491}{514800}$ & $\frac{134401437}{11211200}$ \\\hline	
6	&0 & 0 & $\frac{3125}{5544}$ & $\frac{1831963}{356400}$ & $\frac{18101799}{2094400}$ \\\hline	
8	&0 & 0 & 0 & $\frac{823543}{2316600}$ & $\frac{17891847}{5532800}$ \\\hline	
10	& 0 & 0 & 0 & 0 & $\frac{43046721}{206926720}$ \\\hline	
\end{tabular} \caption{Numerical value of $g_{nl}^2$. } \label{table1}
\end{table}

\medskip
To summarize,
$a_{2,1}$ and therefore the Wilson coefficient $\beta$ of the six derivative operator $(\nabla_\mu\partial_\nu\phi)^2(\partial_\rho\phi)^2$
vanish in the open superstring, which provides a concrete example for UV completion with higher spins and a non-positive $\beta$.

\subsection{KK graviton }\label{kk}

So far we have discussed the spin-dependence of the coefficients of the six-derivative operator based on the Froissart-Martin type bound on top of the unitarity and analyticity of scattering amplitudes. An important assumption there was that gravity is subdominant as we noted below Eq.~\eqref{SA0residue}. Even before looking at the six-derivative operator, it is known that the positivity of the $s^2$ coefficient is subtle in the presence of gravity because the $t$-channel graviton exchange diagram behaves as $\displaystyle\sim\frac{s^2}{t}$ in the forward limit $t\to0$ and dominates over contributions from massive states~\cite{Adams:2006sv}. Similarly, at least naively, it is not clear if the spin-dependence of the $s^2t$ coefficient discussed in the previous subsections is applicable in the presence of gravity\footnote{
\label{positivity_gravity}
Recently Ref.~\cite{Hamada:2018dde} demonstrated that the positivity of the $s^2$ coefficient holds if the contributions to the $s^2$ coefficient from Regge states (which UV complete gravity) are subdominant compared to those from other massive states. This provides a quantitative criterion for the statement that gravity is negligible. A similar argument should hold for the $\mathcal{O}(s^2t^n)$ coefficients.}. Therefore, it is nontrivial if the sign of the six-derivative operator can be used to probe spins of heavy states in models with the Kaluza-Klein (KK) graviton, where the massless graviton inevitably appears. For this reason, in this subsection we study the six-derivative operator generated by the KK graviton by evaluating the four-point scattering amplitude explicitly. 

\medskip
What characterizes the KK graviton $\chi_{\mu\nu}$ is its coupling to the energy-momentum tensor $T_{\mu\nu}$ of other particles. Its 4D Lagrangian reads
\begin{align}
\mathcal{L}_\chi=-\frac{1}{4}\chi^{\mu\nu}\mathcal{E}_{\mu\nu}^{\alpha\beta}\chi_{\alpha\beta}-\frac{m^2}{8}\left(\chi_{\mu\nu}^2-\chi^2\right)
+g\chi^{\mu\nu}T_{\mu\nu}+\ldots\,,\label{kklag}
\end{align}
where we introduced $\chi=\chi^\mu_\mu$ and $g$ parametrizes the coupling of the KK graviton and the energy-momentum tensor. The dots stand for interaction terms with two or more $\chi$, which are not relevant for our purpose. The kinetic operator $\mathcal{E}_{\mu\nu}^{\alpha\beta}$ is defined by
\begin{align}
\mathcal{E}_{\mu\nu}^{\alpha\beta}\chi_{\alpha\beta}
=-\frac{1}{2}
\bigg[
\Box \chi_{\mu\nu}-\partial_\mu\partial_\alpha \chi^\alpha_\nu
-\partial_\nu\partial_\alpha \chi^\alpha_\mu
+\partial_\mu\partial_\nu \chi
-\eta_{\mu\nu}\left(\Box \chi-\partial_\alpha\partial_\beta \chi^{\alpha\beta}\right)
\bigg]\,.
\end{align}
Note that the kinetic and mass terms are the standard Fierz-Pauli ones (see, e.g.,~\cite{Han:1998sg,Hinterbichler:2011tt,deRham:2014zqa}). This type of Lagrangian appears, e.g., in the Randall-Sundrum (RS) I scenario~\cite{Randall:1999ee}, which has a discrete spectrum of the KK gravitons. Inflaton in this context has been studied in~\cite{Kaloper:1998sw,Kaloper:1999sm,Nihei:1999mt,Kim:1999ja,Lukas:1999yn,Maartens:1999hf,Giudice:2002vh,Im:2017eju} and the oscillatory features of non-Gaussianities associated with the on-shell KK graviton creation were studied recently in~\cite{Kumar:2018jxz}. We emphasize that the mass $m$ and the coupling $g$ are model-dependent, but the existence of the interaction $\chi^{\mu\nu}T_{\mu\nu}$ is universal due to the KK graviton nature.

\medskip
We then study the effective interactions of massless scalars mediated by the KK graviton. For this, we use the energy-momentum tensor,
\begin{align}
T_{\mu\nu}=\partial_\mu \phi \partial_\nu \phi -  \frac{1}{2} \eta_{\mu\nu} \partial_\sigma \phi \partial^\sigma \phi\,,
\end{align}
where we dropped terms arising from interaction terms because they are not relevant for our argument based on four-point amplitudes. Recall that the propagator of the KK graviton is given by (see, e.g.,~\cite{Han:1998sg,Hinterbichler:2011tt,deRham:2014zqa})
\begin{align}
P_{\mu\nu ab}=\frac{F_{\mu\nu ab}}{m^2+k^2}~,
\end{align}
with the projector,
\begin{align}\nonumber
F_{\mu\nu a b} &=
\eta_{\mu(a} \eta_{\nu b)}  - \frac{1}{3} \eta_{\mu\nu} \eta_{ab}
\\
&\quad
+ \frac{1}{m^2} \left[ k_{a} k_{(\mu} \eta_{\nu)b} + k_b k_{(\mu} \eta_{\nu)a}  -\frac{1}{3} k_\mu k_\nu \eta_{ab} - \frac{1}{3} k_a k_b \eta_{\mu\nu} \right]
+\frac{2}{3}\frac{k_{\mu}k_{\nu}k_{a}k_{b}}{m^4 }
\,.
\end{align}
Here we used a normalized symmetrizer $\displaystyle A_{(ab)}=\frac{1}{2}(A_{ab}+A_{ba})$. We then obtain the four-point amplitude of identical massless scalars,
\begin{align}
M(s,t)  = \frac{g^2}{6} \left[
\frac{-2 s^2+3 \left(t^2+u^2\right)}{m^2-s}+
\frac{-2 t^2+3 \left(u^2+s^2\right)}{m^2-t}+
\frac{-2 u^2+3 \left(s^2+t^2\right)}{m^2-u}
\right]\,,
\end{align}
where $u=-(s+t)$. 
Note that the amplitude behaves as $\sim s^2$ for large $s$, so that it requires an appropriate UV completion at some scale.
On the other hand, the IR expansion reads
\begin{align}
\label{MKK_IR}
M(s,t) = \frac{4g^2}{3m^2} \left(s^2+st+t^2\right) +\frac{5g^2}{2m^4} \left(s^2 t+st^2\right)+\cdots\,.
\end{align}
We find that both of $s^2$ and $s^2 t$ have a positive coefficient. Interestingly, the sign happens to be the same as the massive spin $2$ exchange discussed in Sec.~\ref{beyond}. Therefore, a non-positive coefficient $\beta$ implies a spinning heavy state also in the KK graviton case. We again emphasize that the previous argument cannot directly be applied to the KK graviton case at least naively because of the $t$-channel graviton pole mentioned earlier, so that a separate analysis was required\footnote{Indeed, the IR coefficients in Eq.~\eqref{MKK_IR} are different from the ones in Eqs.~\eqref{a_p,0}-\eqref{a_p,1}, even though the signs are the same. It does not imply any contradiction in general, but it could provide an interesting direction of the future studies:
As we mentioned at footnote~\ref{positivity_gravity}, we expect that the spin-dependence~\eqref{a_p,1} of the six-derivative term will hold even in the presence of gravity, if the contribution of the Regge states is subdominant.
If it is true, it would imply that the general results should hold if the KK graviton coupling $g$ in Eq.~\eqref{kklag} is big enough (or if the KK scale is low enough) to dominate over the Regge states effect.
Then, the Lagrangian~\eqref{kklag} with a big $g$ (or a small $m$), which is phenomenologically interesting, could be incomplete even as a low-energy EFT because of the mismatch. It would require other couplings such as the dilatonic coupling to resolve the mismatch for example.
If the mismatch could never be resolved by such couplings, it would imply that models with the KK graviton with a big $g$ (compared to the gravitational coupling) and a small $m$ (compared to the mass scale of Regge states, i.e., the string scale in stringy UV completion) are in the swampland. We leave such a direction for the future work.
}.

\bigskip
To conclude this section, a nonpositive coefficient $\beta$ of the six derivative operator $(\nabla_\mu\partial_\nu\phi)^2(\partial_\rho\phi)^2$ is a signature of intermediate spinning particles with even spin $l=2,4,...$. This conclusion is applicable to both (a) non-gravitational theories which respect the Froissart-Martin bound and (b) models with the KK graviton. See also Table~\ref{table:summary} for a summary of this section.

\section{De Sitter Four-Point Functions }\label{4pt}

In the rest of the paper we apply our argument on sign of the six derivative operator to cosmological settings. We start from the effective action of a massless scalar $\phi$ with a shift symmetry on exact de Sitter space:
\begin{align}\label{theaction}
S =\int d\tau d^3 x  \sqrt{-g} \bigg[ - \frac{1}{2} (\partial_\mu \phi)^2 + \frac{\alpha}{\Lambda^4} (\partial_{\mu}\phi\partial^\mu\phi)^2 + \frac{\beta}{\Lambda^6} (\nabla_\mu\partial_\nu\phi)^2 (\partial_\rho \phi)^2+\ldots\bigg]\,,
\end{align}  
where the dots stand for operators with more derivatives and/or $\phi$. As we have discussed, the four-derivative operator has to be always positive $\alpha>0$, whereas the sign of $\beta$ tells us which of intermediate heavy scalars and spinning states are dominant. Here one might wonder if we can apply our argument based on the flat space scattering directly to the de Sitter case. Indeed, the counting of derivatives becomes ambiguous due to curvature couplings. For example, we may write a six-derivative operator,
\begin{align}
\frac{\tilde{\beta}}{\Lambda^6}R(\partial_\mu\phi\partial^\mu\phi)^2\sim
\frac{H^2}{\Lambda^2}\frac{\tilde{\beta}}{\Lambda^4}(\partial_\mu\phi\partial^\mu\phi)^2\,,
\end{align}
which vanishes on flat space. However, such effects are subdominant to the four-derivative operator $\alpha$ by a factor of $H^2/\Lambda^2$, hence the use of our flat space results is justified as long as $H\ll \Lambda$. It is indeed the case because the de Sitter temperature is around the Hubble scale $H$ and the validity of the EFT description requires $H\ll \Lambda$. 
Note that thermal production of heavy particles are exponentially suppressed by the Boltzmann factor, so that their effects are negligible compared to the above mentioned curvature effect.

\medskip
Our question is now to clarify if we can distinguish the six-derivative operator from the leading-order four-derivative operator. For this purpose, in the rest of this section, we calculate late-time four-point functions generated by these two operators and discuss difference in the shape. Note that the four-point functions calculated in this section are applicable also to inflation at least in the regime where the special conformal symmetry is weakly broken. We also study inflationary three-point functions in the next section.

\subsection{A brief review of the in-in formalism }

We use the in-in formalism to calculate late-time $n$-point functions of $\phi$ in de Sitter space. With an interaction Hamiltonian $H_{I}(\tau)$, the late-time correlator of an operator $O(\tau)$ is evaluated as (see, e.g,~\cite{Weinberg:2005vy,Chen:2010xka,Wang:2013eqj})
\begin{align}\label{expression1}
\langle O(0)\rangle  = \langle0| \Big[ \bar T e^{i \int_{-\infty}^0d \tau_1\, H_{I} (\tau_1)} \Big]\, O(0)\, \Big[ T e^{-i \int_{-\infty}^0d\tau_2\, H_{I} (\tau_2) } \Big] |0\rangle~,
\end{align}
where $T$ and $\bar T$ denote the time-ordering and anti-time-ordering operators, respectively. We choose the Bunch-Davies vacuum as the state $|0\rangle$. For our purpose, the first order in $H_I (\tau)$ is enough to work with, so that we have
\begin{align}\label{expression2}
\langle O(0)\rangle = 2 {\rm Im} \int_{-\infty}^0 d\tau \langle 0|O(0) H_I (\tau) |0\rangle~,
\end{align}
where we assumed $O$ is real. In the conformal time, the second order action reads
\begin{align}
S_2 = \int  d \tau d^3 x\frac{1}{2H^2\tau^2} \Big[\phi'^2 -  (\partial_i \phi)^2 \Big]\,,
\end{align}
where $H$ is the (constant) Hubble parameter and the prime denote a derivative in conformal time $\tau$. In the interaction picture, the massless scalar $\phi$ can then be quantized as
\begin{align}
\phi_{\mathbf k} (\tau) = u_k(\tau) a_{\mathbf k} + u_k^*(\tau) a_{-\mathbf k}^\dagger~,
\end{align}
with the standard commutation relations,
\begin{align}
[a_{\mathbf k},a_{\mathbf k'}^\dagger]=(2\pi)^3\delta^{(3)}({\mathbf k}-{\mathbf k}')\,,
\quad
\text{others}=0\,.
\end{align}
The mode function follows the equation of motion,
\begin{align}
u_k''-2\tau^{-1}u_k'+k^2u_k=0\,,
\end{align}
and it is normalized as
\begin{align}
\left(u_k{u_k^*}'-u_k'u_k^*\right)=iH^2\tau^2\,.
\end{align}
For the Bunch-Davis vacuum, we have a mode function,
\begin{align}
\label{mode}
u_k (\tau) = \frac{H}{\sqrt{2 k^3}} (1+ i k \tau) e^{- i k \tau}\,.
\end{align}

\subsection{Comparison of the two shapes }

We then investigate the shape of four-point functions with the Lagrangian~\eqref{theaction}.
Since there are no three-point interactions, the interaction Hamiltonian relevant for us is simply a minus of the quartic Lagrangian:
\begin{align}
	 H_4
	 &=\int d^3 x\bigg[ -\frac{\alpha}{\Lambda^4} \big(\phi'^2-(\partial_i\phi)^2\big)^2 
	 \nonumber
	 \\
	 &\qquad\qquad\,\,\,
	 +\frac{\beta}{\Lambda^6} H^2\tau^2
\Big(\phi'^2-(\partial_i\phi)^2\Big)
\nonumber
\\
&\qquad\qquad\quad\,\,\,\,
\times\Big((\phi''+\tau^{-1}\phi'\big)^2-2(\partial_i\phi'+\tau^{-1}\partial_i\phi)^2+(\partial_i\partial_j\phi+\tau^{-1}\delta_{ij}\phi')^2\Big)
	 \bigg]
	 ~,
\end{align} 
where the subscript $4$ indicates that it contains four $\phi$.
It is straightforward to calculate the four-point function of $\phi$ by using Eq.~\eqref{expression2}. Schematically, we write
\begin{align}
& \langle \phi_{\mathbf k_1}(0) \phi_{\mathbf k_2}(0) \phi_{\mathbf k_3}(0) \phi_{\mathbf k_4} (0)\rangle=(2\pi)^3\delta(
\sum_i\mathbf k_i)
\Big[
\alpha \mathcal{A}(\mathbf k_i)+\beta \mathcal{B}(\mathbf k_i)
\Big]\,,
\end{align}
where $\mathcal{A}(\mathbf k_i)$ and $\mathcal{B}(\mathbf k_i)$ are associated with the four-derivative and six-derivative operators, respectively. Since the full shape is somewhat complicated to present, our shape analysis focuses on two particular limits of momentum configurations, the equilateral limit $ k_1 =  k_2= k_3= k_4$ and the flat space limit $k_{1234}=k_1+k_2+k_3+k_4\to 0$, which are useful to distinguish the two shapes.

\begin{figure}[htbp] 
	\centering 
	\includegraphics[width=8.5cm]{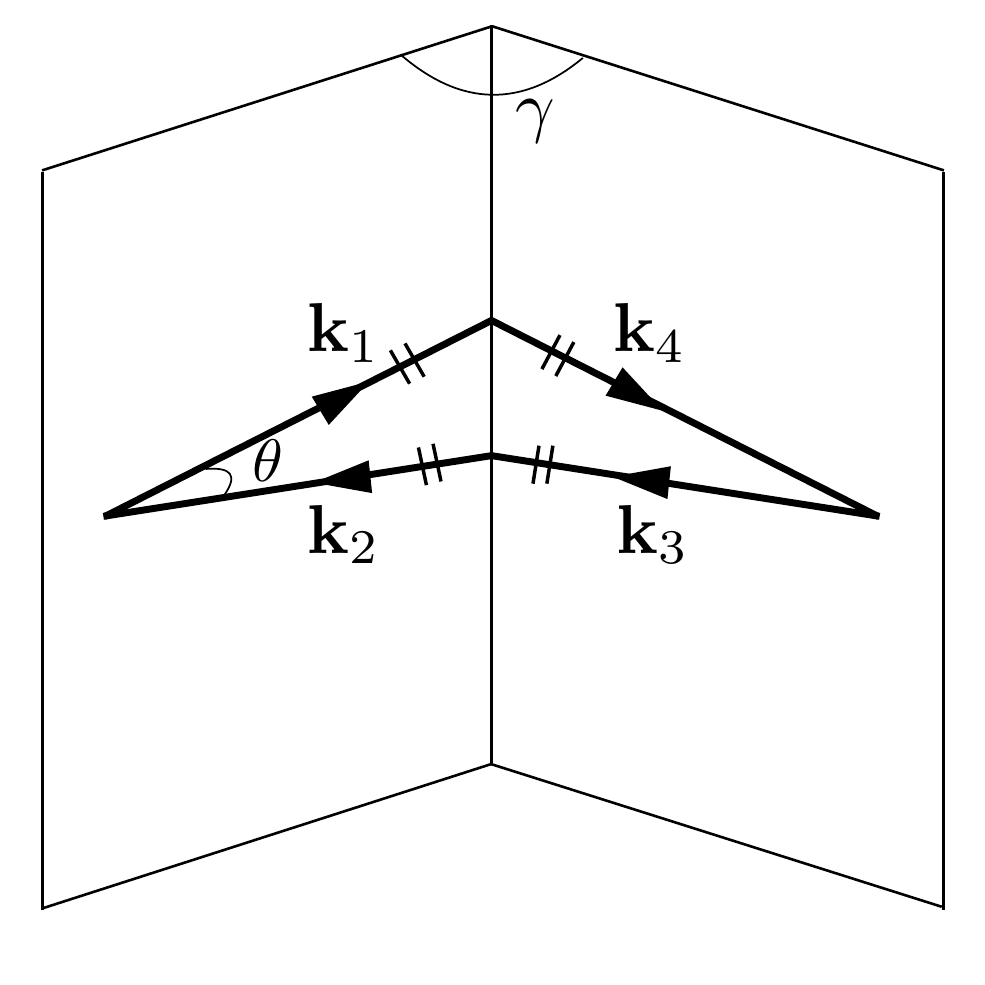}   
	\caption{Equilateral configurations of four momenta
	} \label{tranglefonfiguration}
\end{figure}

\paragraph{ Equilateral limit}

As depicted in Fig.~\ref{tranglefonfiguration}, momentum configurations in the equilateral limit are characterized by $k_1=k_2=k_3=k_4=k$, the angle $\gamma$ between the two isosceles triangles, and the angle $\theta$ between $\mathbf k_1$ and $-\mathbf k_2$ up to an overall rotation which does not affect scalar four-point functions. In this language we have
\begin{align}\nonumber
 &\mathbf k_1\cdot \mathbf k_2 = -k^2 \cos\theta\,, \quad \mathbf k_1\cdot \mathbf k_3=\mathbf k_2\cdot \mathbf k_4
=k^2\left(\frac{\cos\theta+1}{2}\cos\gamma+\frac{\cos\theta-1}{2}\right)
\,,\\
 &\mathbf k_3\cdot \mathbf k_4 = - k^2 \cos\theta\,,\quad
\mathbf k_1\cdot \mathbf k_4=\mathbf k_2\cdot \mathbf k_3
=k^2
\left(-\frac{\cos\theta+1}{2}\cos\gamma+\frac{\cos\theta-1}{2}\right)\,.
\end{align}  
Then, the shape functions $\mathcal{A}(\mathbf k_i)$ and $\mathcal{B}(\mathbf k_i)$ are evaluated as
\begin{align}
 \mathcal{A}(\mathbf k_i) &
= \frac{H^8}{256\Lambda^4k^9}\big(\cos^2\theta(103\cos^2\gamma+309  )-206\cos\theta\sin^2\gamma +103\cos^2\gamma+173 \big)\,,
\\
\nonumber
\mathcal{B}(\mathbf k_i)
&=\frac{H^{10}}{4096\Lambda^6k^9}\big(- 1311\cos^3\theta\sin^2\gamma-\cos^2\theta(1111\cos^2\gamma+8577)
\\
&\qquad\qquad\qquad\,
+6155\cos\theta\sin^2\gamma-(3733\cos^2\gamma+3155)   \big)\,.
\end{align}
We find that the $\theta$-dependence is qualitatively different among the two: the four-derivative and six-derivative operators generate up to the second and the third order harmonics in $\theta$, respectively. Therefore, the coefficient of $\cos 3\theta$ can be used to probe the six-derivative operator without swamped by the four-derivative one.

\medskip
More generally, the $n$-th order harmonics in $\theta$ is sensitive to the $2n$-derivative operators. It is essentially because mode functions of all external scalars and their conformal time derivatives do not depend on the angles $\theta$ and $\gamma$, and thus all the angular dependence is through spatial derivatives in the effective interactions. In this way, the equilateral configurations parameterized by $\theta$ and $\gamma$, and especially the $\theta$-dependence of four-point functions is useful to separate out contributions from each order in the derivative expansion.

\paragraph{ Flat space limit $k_{1234}\to 0$} 

Another interesting limit is the so-called flat space limit $k_{1234}=k_1+k_2+k_3+k_4\to 0$ with $k_i\neq0$, which is analogous to the flat space limit of AdS correlators proposed by Raju~\cite{Raju:2012zr}. Note that this limit cannot be achieved by physical momentum configurations of dS correlators, hence it requires analytic continuation. By virtue of analytic continuation, we may probe deep inside the horizon $\tau\rightarrow -\infty$ or in other words high-energy scattering on de Sitter. As demonstrated in~\cite{Arkani-Hamed:2015bza,Arkani-Hamed:2017fdk}, correlators in this limit are proportional to the corresponding flat space amplitudes.
In our setup the flat space limit $k_{1234}\rightarrow 0$ reads
\begin{align}\label{flat4d}
\mathcal{A}(\mathbf k_i) &=24 \frac{ \alpha}{\Lambda^4} \frac{H^8 (k_{12}^2-k_I^2)(k_{34}^2-k_I^2)}{4 k_1^2 k_2^2 k_3^2 k_4^2 k_{1234}^5 } + {\rm 2 \,\, Permutations} + \mathcal O(k_{1234}^{-4})~, 
\\ \label{flat6d}
\mathcal{B}(\mathbf k_i) &= -180 \frac{\beta}{\Lambda^6} \frac{ H^{10} (k_{12}^2-k_I^2)^2 (k_{34}^2 - k_I^2) }{4 k_1^2 k_2^2 k_3^2 k_4^2 k_{1234}^7} + {\rm 5 \,\, Permutations} + \mathcal O(k_{1234}^{-6})~,
\end{align}
where we used shorthand notations $k_{12}\equiv k_1+k_2$, $k_{34}\equiv k_3+k_4$, and $\mathbf k_I = \mathbf k_1+\mathbf k_2$. 
Note that the leading order of $\mathcal{A}$ gives the flat space scattering amplitude $\sim s^2$, if we identify $k_{12}$ and $k_{34}$ as the energy variable and $k_I$ as the three momentum.  
On the other hands, $\mathcal{B}$ gives the flat space scattering amplitude $\sim s^3$. More detailed analysis of the flat space limit is given in Appendix~\ref{app:flat}, to which we refer the readers with interests.

\section{Inflationary Three-point Functions }\label{3pt}
Next we investigate inflationary three-point functions.
We use the same inflaton Lagrangian as Sec.~\ref{4pt} (with an appropriate slow-roll potential) and turn on a time-dependent inflaton background $\bar{\phi}(t)$ to write
\begin{align}
\phi(t, \mathbf x) = \bar{\phi}(t)+\delta \phi (t, \mathbf x)\,,
\end{align}
where
$\delta\phi(t,{\mathbf x})$ denotes inflaton fluctuations. We assume that $\dot{\bar{\phi}}$ is nearly constant under the slow-roll approximation (the dot denotes a derivative in physical time $t$ as usual). Also, as we mentioned earlier, we focus on the regime the de Sitter conformal symmetry is weakly broken, so that we neglect higher order terms in $\dot{\bar{\phi}}$ in the following analysis. Under this assumption, the cubic interaction Hamiltonian is given by
\begin{align}
	 H_3
	 &=\int d^3 x\bigg[ \frac{4\alpha}{\Lambda^4} \frac{1}{H\tau}\dot{\bp}\dep'(\dep'^2-(\partial_i\dep)^2) 
	 \nonumber
	 \\
	 \nonumber
	 &\qquad\qquad\,\,
	 -\frac{2H\beta\dot{\bp}}{\Lambda^6} \Big[
	\tau\delta\phi'
	\Big((\dep''+\tau^{-1}\dep'\big)^2-2(\partial_i\dep'+\tau^{-1}\partial_i\dep)^2+(\partial_i\partial_j\dep+\tau^{-1}\delta_{ij}\dep')^2\Big)
	\\
	&\qquad\qquad\qquad\qquad\,\,\,
	+\Big(\dep'^2-(\partial_i\dep)^2\Big)
	\Big(\partial_i^2\dep+3\tau^{-1}\dep'\Big)
	 \Big]\bigg]
	 ~,
	 \label{3pt_Ham}
\end{align} 
where the subscript 3 indicates that it contains three $\delta \phi$. Since the cubic interaction is $\mathcal{O}(\dot{\bar{\phi}})$, corrections to the quadratic Lagrangian are not relevant for our purpose, so that we use the uncorrected linear equation of motion,
\begin{align}
\delta \phi''-2\tau^{-1}\delta\phi'-\partial_i^2\delta\phi=0\,,
\label{eom_dp}
\end{align}
and the corresponding mode functions for canonical quantization.
When calculating three-point functions, we may use Eq.~\eqref{eom_dp} to simplify the cubic Hamiltonian~\eqref{3pt_Ham} as
\begin{align}
\label{H_3_simplified}
	 H_3&=\int d^3 x\bigg[ \frac{4\alpha\dot{\bp}}{H\Lambda^4\tau} \dep'(\dep'^2-(\partial_i\dep)^2) 
	 -\frac{4H\beta\dot{\bp}}{\Lambda^6\tau} \Big[
\dep'\Big(\dep'^2-(\partial_i\dep)^2\Big)
	+2(\dep')^3 \Big]\nonumber\\
	&\qquad\qquad   - \frac{H\beta\dot{\bp}}{\Lambda^6}\frac{d}{d\tau}\Big(3\delta\phi'^3 + 2\tau\delta\phi'^2(\partial_i^2\delta\phi) - 3\delta\phi'(\partial_i\delta\phi)^2 + \tau\partial_i^2\delta\phi(\partial_j\delta\phi)^2 \Big)\bigg]
	 ~,
\end{align}
where the second line is a total time derivative that is not relevant in the following calculation\footnote{
In the in-in formalism, it is sometimes dangerous to neglect total time derivatives because they may provide a non-vanishing boundary term in general. However, it is easy to see that the boundary terms we dropped vanish at the future boundary $\tau=0$ and thus we may safely focus on the first line of Eq.~\eqref{H_3_simplified}.
}.
Note that the first term of the $\beta$ term in the first line is the same as the $\alpha$ one up to an overall constant.
Now it is straightforward to calculate three-point functions of the scalar curvature perturbation $\zeta=-(H\delta\phi)/\dot{\bar{\phi}}$. We schematically write
\begin{align}
\nonumber
\lran{\zeta_{\bfk_1} \zeta_{\bfk_2} \zeta_{\bfk_3}} &= -\frac{H^3}{\dot{\bar\phi}^3} \lran{\delta\phi_{\bfk_1} \delta\phi_{\bfk_2} \delta\phi_{\bfk_3}} 
\\
&\equiv (2\pi)^7 P_\zeta^2 \frac{1}{k_1^2k_2^2k_3^2} \delta(\sum_i\mathbf k_i)\Big[
\alpha \mathcal{A}_{\rm 3pt}(\mathbf k_i)+\beta \mathcal{B}_{\rm 3pt}(\mathbf k_i)
\Big]\,,
\label{3pt_zeta}
\end{align}
where $\mathcal{A}_{\rm 3pt}$ and $\mathcal{B}_{\rm 3pt}$ are contributions from the four-derivative and six-derivative operators, respectively. 
$P_\zeta$ is the scalar power spectrum given by
\begin{align}
P_\zeta=\frac{H^2}{(2\pi)^2}\frac{H^2}{\dot{\bp}^2}\,.
\end{align}
Then, the shape function $\mathcal{A}_{\rm 3pt}$ is evaluated as~\cite{Creminelli:2003iq}
\begin{align}
\nonumber
\mathcal{A}_{\rm 3pt}(k_1,k_2,k_3)=&\frac{\dot{\bp}^2}{\Lambda^4k_1k_2k_3(k_1+k_2+k_3)^2}
\\
\nonumber
&\times\big[ k_1^5+2k_1^4k_2+2k_1^4k_3-3k_1^3k_2^2+2k_1^3k_2k_3-3k_1^3k_3^2-3k_1^2k_2^3
\\
\nonumber
&\quad-8k_1^2k_2^2k_3-8k_1^2k_2k_3^2-3k_1^2k_3^3+2k_1k_2^4+2k_1k_2^3k_3-8k_1k_2^3k_3^2+2k_1k_2k_3^3
\\
&\quad+2k_1k_3^4+k_2^5+2k_2^4k_3-3k_2^3k_3^2-3k_2^2k_3^3+2k_2k_3^4+k_3^5\big]~.
\end{align}
On the other hand, $\mathcal{B}_{\rm 3pt}$ is
\begin{align}
\mathcal{B}_{\rm 3pt}(k_1,k_2,k_3)=-\frac{H^2}{\Lambda^2}\mathcal{A}_{\rm 3pt}(k_1,k_2,k_3)+\mathcal{B}'_{\rm 3pt}(k_1,k_2,k_3)
\end{align}
with $\mathcal{B}'_{\rm 3pt}$ defined by 
\begin{align}
\mathcal{B}'_{\rm 3pt}(k_1,k_2,k_3)=\frac{24\dot{\bp}^2H^2k_1k_2k_3}{\Lambda^6(k_1+k_2+k_3)^3}\,,
\end{align}
which is associated to the second term $(\delta \phi')^3$ of the $\beta$ term in Eq.~\eqref{H_3_simplified}.

\begin{figure}[t]
 \begin{minipage}{0.5\hsize}
  \begin{center}
\includegraphics[width=70mm]{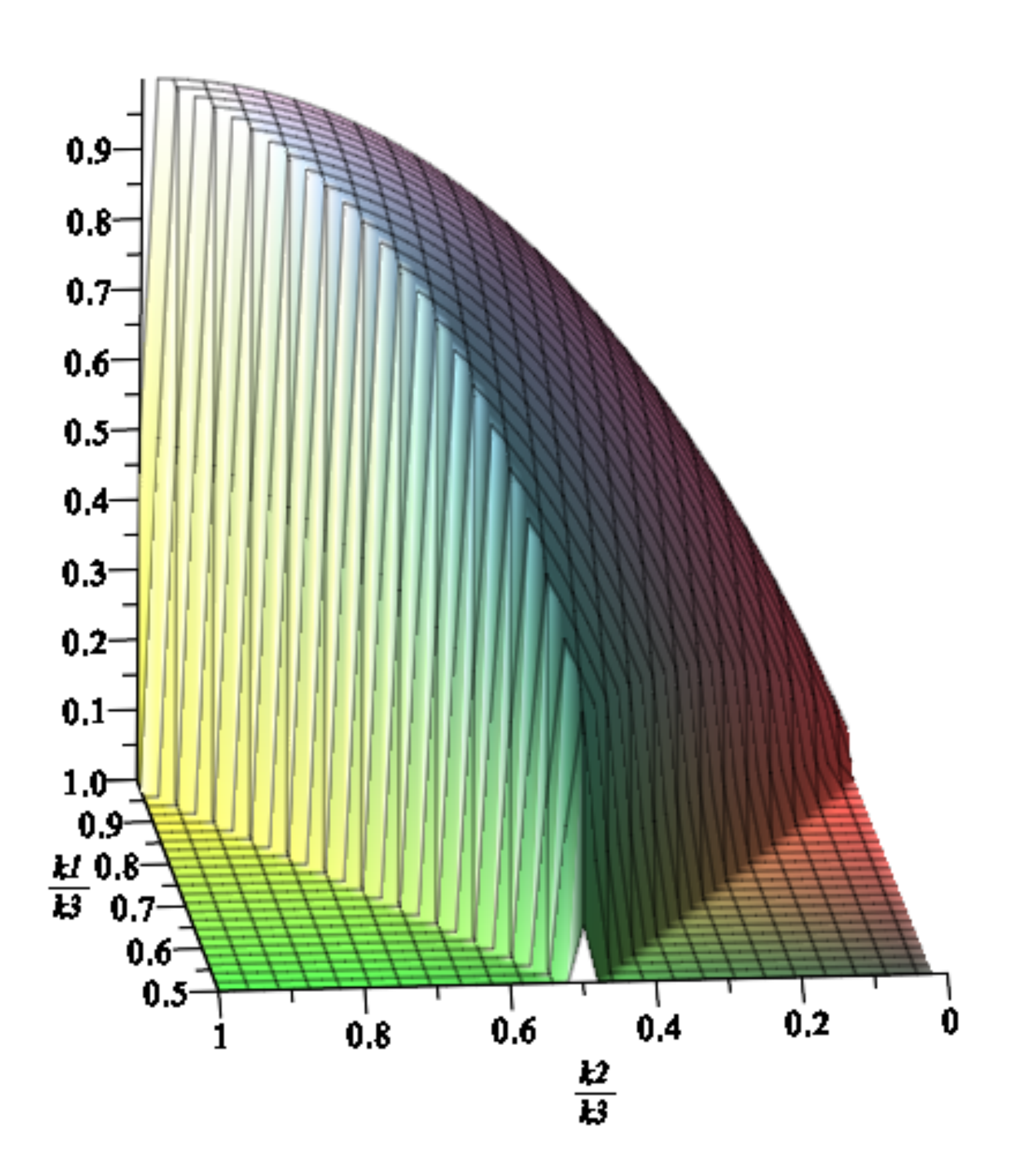}
  \end{center}
 \end{minipage}
 \begin{minipage}{0.5\hsize}
  \begin{center}
  	\includegraphics[width=70mm]{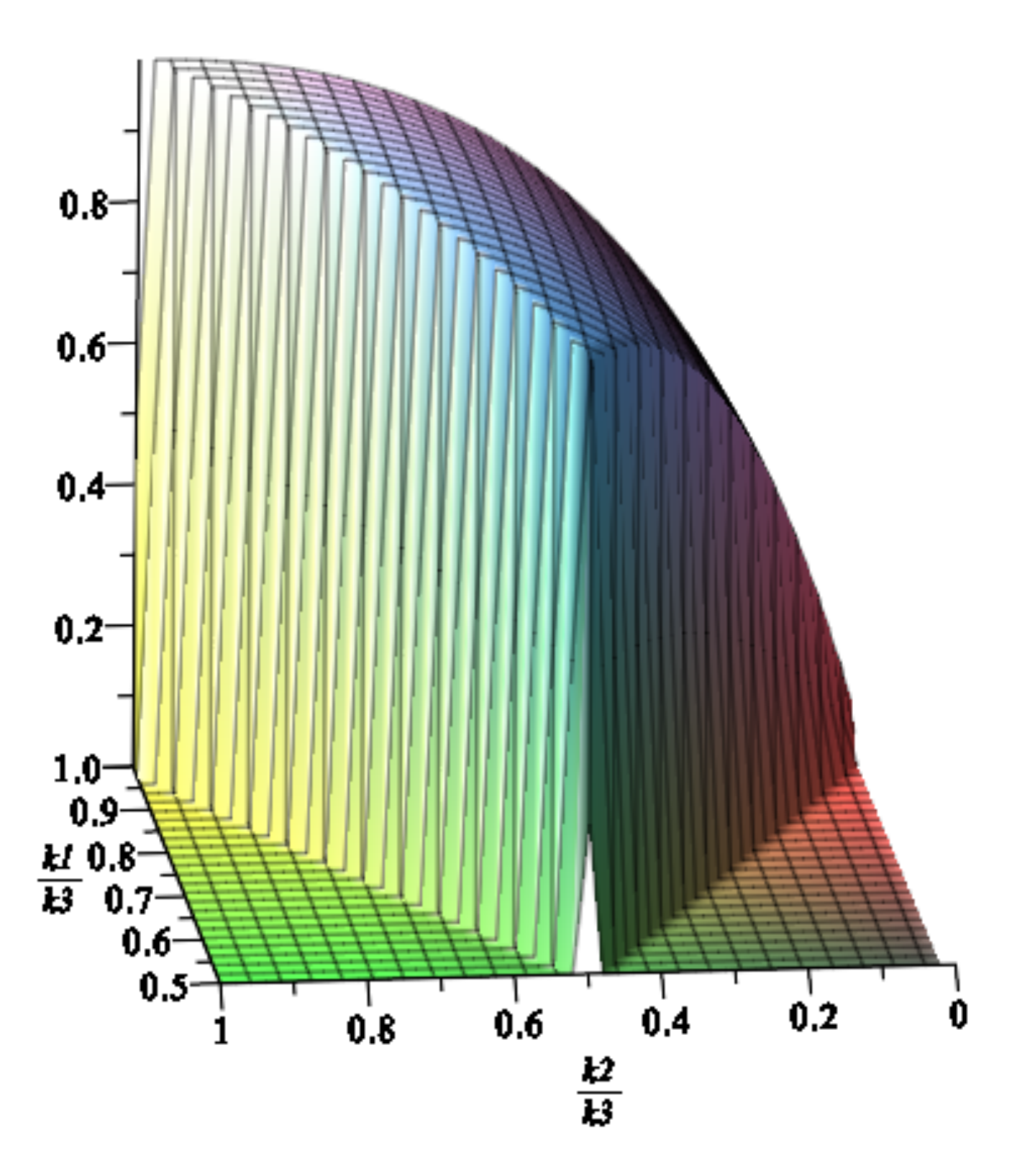}  
\end{center}
 \end{minipage}
\caption{
Shape of $\mathcal{A}_{\rm 3pt}$(left) and $\mathcal{B}'_{\rm 3pt}$(right):
Both of them have a peak at the equilateral configuration $k_1=k_2=k_3$, but the height at the folded configuration $k_1=k_2=k_3/2$ is different by a factor $\sim3$.
 }
 \label{nG}
\end{figure}

\medskip
Next we discuss how to distinguish the three-point functions generated by the six-derivative operator from other contributions.
First, as depicted in Fig.~\ref{nG}, both the leading contribution from the four-derivative operator and the next-to-leading one from the six-derivative operator have a peak at the equilateral configuration. Therefore we need to look at details of the shape.
For this purpose, we may use, e.g., the ratio between the equilateral configuration and the folded one:
\begin{align}
\frac{\mathcal{A}_{\rm 3pt}(0.5,0.5,1)}{\mathcal{A}_{\rm 3pt}(1,1,1)}\sim0.32\,, \qquad\qquad\qquad
\frac{\mathcal{B}'_{\rm 3pt}(0.5,0.5,1)}{\mathcal{B}'_{\rm 3pt}(1,1,1)}\sim0.84\,.
\end{align}
It is also worth clarifying if the shape has degeneracy with other operators we have neglected. As we mentioned, the shape $\mathcal{B}'_{\rm 3pt}$ is from the operator $(\delta \phi^{\prime})^3$, which arises from the six-point effective interaction $(\partial_\mu\phi \partial^\mu\phi)^3$ as well. One would then wonder if $(\nabla_\mu\partial_\nu\phi)^2 (\partial_\rho \phi)^2$ and $(\partial_\mu\phi \partial^\mu\phi)^3$ are degenerate as long as we look at inflationary three-point functions. However, recall that $(\partial_\mu\phi \partial^\mu\phi)^3$ provides the interaction $(\delta \phi^{\prime})^3$ at the cubic order in $\dot{\bar{\phi}}$. To work at this order, we need to take into account corrections to the linear equation of motion~\eqref{eom_dp} when simplifying the cubic Hamiltonian, which gives a new operator on top of the ones in Eq.~\eqref{H_3_simplified}. Hence, the potential degeneracy of $(\nabla_\mu\partial_\nu\phi)^2 (\partial_\rho \phi)^2$ and $(\partial_\mu\phi \partial^\mu\phi)^3$ are resolved by carefully studying higher order terms in $\dot{\bar{\phi}}$ which we neglected. Besides, these two operators generate different shapes of inflationary four-point functions because the four-point interaction of $\delta \phi$ originating from $(\nabla_\mu\partial_\nu\phi)^2 (\partial_\rho \phi)^2$ is Lorentz invariant, but the one from $(\partial_\mu\phi)^6$ is not.

\medskip
To end this section, we estimate the nonlinearity parameter $f_{NL}$ and note the limitation of our present work for the near future phenomenology. First, $f_{NL}$ sourced by the four-derivative operator is estimated as
\begin{align}
f_{NL}\sim\frac{\lran{\zeta\zeta\zeta}}{\lran{\zeta\zeta}\lran{\zeta\zeta}}\sim\frac{\dot{\bp}^2}{\Lambda^4}\,,
\end{align}
so that the observable non-Gaussianity $f_{NL}\gtrsim1$ can be achieved only when $\dot{\bp}^2\gtrsim\Lambda^4$. However, contributions from the six-point interaction $\displaystyle\frac{\gamma}{\Lambda^{8}}(\partial_\mu\phi\partial^\mu\phi)^3$ and higher are non-negligible in the regime $\dot{\bp}^2\gtrsim\Lambda^4$ as long as we employ the ordinary order-estimation of the effective theory, i.e., if the dimensionless parameter $\gamma$ is $\mathcal{O}(1)$ for example. Therefore, our argument based on the truncation at the four-point level is not applicable to the regime $f_{NL}\gtrsim1$ generically, which motivates further studies along the line of our present work.
\section{Outlook }
\label{conclusion}

To enlarge the scope of the cosmological collider, we initiated a program decoding imprints of heavy particles from effective interactions of primordial perturbations. In this paper, as a first step in  this direction, we studied spin-dependence of four-point effective interactions of the inflaton. It is well known that the sign of the four-derivative operator $(\partial_{\mu}\phi\partial^\mu\phi)^2 $ is universally positive~\cite{Adams:2006sv} as a consequence of unitarity and analyticity of scattering amplitudes together with the Froissart-Martin bound. In contrast to the universal positivity bound, we demonstrated that the sign of the six-derivative operator $(\nabla_\mu\partial_\nu\phi)^2 (\partial_\rho \phi)^2$ is positive for intermediate scalars, whereas it is negative for intermediate spinning states.
In particular, a non-positive coefficient requires heavy spinning particles with the spin $s=2,4,...$ in tree-level UV completion.
This result applies, e.g., to (1) non-gravitational theories enjoying the Froissart-Martin bound and (2) effective interactions generated by the KK graviton. We thus conclude that we may probe spin of heavy intermediate states from the sign of effective interactions by going beyond the positivity bound.

\medskip
We also studied phenomenology of primordial non-Gaussianities thereof. Since the six-derivative operator is the next-to-leading order correction in the inflaton effective action, we need to explore how to distinguish it from the leading order correction, i.e., the four-derivative operator. First, we found that they are distinguishable from angular dependence of de Sitter four-point functions, especially in the equilateral limit. On the other hand, both operators generate three-point functions with a peak at the equilateral configuration, so that a detailed analysis of the full shape is required to distinguish the two from three-point functions. We again emphasize that the signals we studied are generated by intermediate off-shell heavy particles, so that it is free from the exponential Boltzmann suppression in contrast to the previous studies in the cosmological collider program~\cite{Chen:2009zp,Baumann:2011nk,Noumi:2012vr,Arkani-Hamed:2015bza}.

\medskip
There are various directions to explore along the line of our present work. First of all, the present paper focused on a regime where the de Sitter conformal symmetry is weakly broken to neglect six-point interactions such as $(\partial_\mu \phi\partial^\mu \phi)^3$ and higher. As we mentioned, primordial non-Gaussianities in this regime are generically small $f_{NL}\lesssim1$ if we employ the usual order-estimation of the effective theory. It would be interesting to explore a concrete UV model which has an observable non-Gaussianity $f_{NL}\gtrsim1$, but negligible higher-point effective interactions. A more challenging, but important direction is to extend our analysis to the effective field theory of inflation~\cite{Cheung:2007st}, or equivalently to incorporate the higher-point inflaton effective interactions, which covers the regime  $f_{NL}\gtrsim1$ in a general context. In this regime, we cannot utilize, e.g., the Lorentz invariance and crossing symmetry anymore (see~\cite{Baumann:2015nta} for related discussions), so that we would need another UV input such as Lorentz symmetry restoration at high energy~\cite{progress}. It would also be interesting to generalize our work to include external spinning particles, e.g., for applications to graviton non-Gaussianities and phenomena other than inflation. We hope to report our progress in these directions elsewhere in the near future.

\section*{Acknowledgements}
We would like to thank Yu-tin Huang, Pablo Soler, Toshiaki Takeuchi, Henry Tye, Yi Wang and Shuang-Yong Zhou for useful discussion and comments.
S.K. is supported in part by the Senshu Scholarship Foundation.
T.N. is supported in part by JSPS KAKENHI Grant Numbers JP17H02894 and JP18K13539, and MEXT KAKENHI Grant Number JP18H04352. SZ is supported in part by ECS Grant 26300316 and GRF Grant 16301917 and 16304418 from the Research Grants Council of Hong Kong.

\appendix
\section{Comments on loops }\label{loopdiagramappendix}

As we mentioned, the intermediate states generating effective interactions can be multi-particle states associated with loops. It will then be useful to demonstrate which sign of $s^2t$ appears in typical loop diagrams. For illustration, here we consider one-loop amplitudes with an internal massive scalar/fermion.

\subsection{Scalar loop } 

Let us begin by the following two-scalar model:
\begin{align}
\mathcal{L}=-\frac{1}{2}(\partial_\mu\phi)^2-\frac{1}{2}(\partial_\mu\sigma)^2-\frac{1}{2}m^2\sigma^2
-\frac{g_1}{2}\phi\sigma^2-\frac{g_2}{4}\phi^2\sigma^2\,,
\end{align}
where $\sigma$ is the massive particle to be integrated out and the interactions are parameterized by the couplings $g_1$ and $g_2$. 
Four-point scattering amplitudes of $\phi$ at one-loop are then
\begin{align}
\nonumber
M(s,t)&=\frac{g_2^2}{2}\Big[
I_{\rm bub}(k_1+k_2)
+I_{\rm bub}(k_1+k_3)
+I_{\rm bub}(k_1+k_4)
\Big]
\\
\nonumber
&\quad
-g_1^2g_2\Big[
I_{\rm tri}(k_1+k_2,-k_4)
+\text{5 permutations }
\Big]
\\
&\quad
+g_1^4\Big[
I_{\rm box}(k_1,k_1+k_2,-k_4)
+\text{2 permutations }\Big]\,.
\end{align}
Here we defined the bubble, triangle, and box integrals by
\begin{align}
\label{bubble_int}
I_{\rm bub} (K_1) & = \int \frac{d^d l}{(2\pi)^d} \frac{1}{(l^2+m^2)((l+K_1)^2+m^2)} \,, 
\\
I_{\rm tri}(K_1,K_2) & = \int   \frac{d^4 l}{(2\pi)^4} \frac{1}{(l^2+m^2)((l+K_1)^2+m^2)((l+K_2)^2+m^2)} \,,
\\
\label{box_int}
I_{\rm box}(K_1,\!K_2,\!K_3) & = \int  \! \frac{d^4 l}{(2\pi)^4} \frac{1}{\!(l^2\!+\!m^2)((l\!+\!K_1)^2\!+\!m^2)((l\!+\!K_2)^2\!+\!m^2)((l\!+\!K_3)^2\!+\!m^2)\!} \,,
\end{align}
where the loop integral over $l$ is already Wick rotated. Also the bubble integral~\eqref{bubble_int} is defined in $d=4-\epsilon$ and the UV divergence $\sim1/\epsilon$ has to be subtracted by a counterterm appropriately. Using the standard Feynman integrals, we may rewrite the integrals as
\begin{align}
I_{\rm bub} (K_1)
& = \frac{\Gamma\left(2-\frac{1}{2} d\right) \Gamma\left(\frac{1}{2} d\right)}{(4 \pi)^{d / 2} \Gamma(2) \Gamma\left(\frac{1}{2} d\right)} \int_0^1 dx  \Big(x(1-x) K_1^2+ m^2\Big)^{-(2-d / 2)} ~,
\\
I_{\rm tri}(K_1,K_2)
& 
 =\frac{1}{16\pi^2} 
  \int_{0}^{1} d x_1
 d x_{2} d x_{3} 
\, \delta\left(x_{1}+x_{2}+x_{3}-1\right)D_{123}^{-1}\,,
\\
I_{\rm box}(K_1,K_2,K_3) 
& =\frac{1}{16\pi^2}  \int_{0}^{1} d x_{1} d x_{2} d x_{3}d x_{4} \delta\left(x_{1}+x_2+x_3+x_{4}-1\right) D_{1234}^{-2}\,,
\end{align}
where we introduced
\begin{align}
D_{123}&= -({K_1} {x_1}+{K_2} {x_2})^2+K_1^2 x_1+K_2^2 {x_2}+m^2\,,
\\
D_{1234}&= -({K_1} {x_1}+{K_2} {x_2}+ {K_3} {x_3})^2+K_1^2 x_1+K_2^2 {x_2}+K_3^2 {x_3}+m^2\,.
\end{align}
It is now easy to evaluate IR coefficients analytically by expanding the integrand in $K_i$. The four- and six-derivative terms of the amplitude then read 
\begin{align}
\nonumber
M(s,t)&=\frac{3 g_1^4 - 8 g_1^2 g_2 m^2 + 6 g_2^2 m^4}{5760\pi^2m^6}\left(s^2+st+t^2\right)
\\
&\quad-\frac{10g_1^4-27g_1^2g_2m^2+18g_2^2m^4}{80640\pi^2m^{10}}st(s+t)\,.
\end{align}
Here and in the next subsection we suppress a constant piece, which can be eliminated by the $\phi^4$ counterterm, as well as higher derivative terms.
We find that the coefficient of $s^2$ is positive as required by unitarity. On the other hand, the sign of $s^2t$ depends on the ratio of $g_1^2$ and $g_2$.

\subsection{Fermion loop }

We next consider a model of a massless (pseudo-)scalar $\phi$ and a fermion $\psi$:
\begin{align}
\mathcal{L}=-\frac{1}{2}(\partial_\mu\phi)^2
+i\bar{\psi}\slash\!\!\!\partial\psi
-m\bar{\psi}\psi
-y\phi\bar{\psi}\Gamma\psi\,,
\end{align}
where $\Gamma=1$ ($\Gamma=i\gamma_5$) when $\phi$ is a scalar (pseudo-scalar).

\paragraph{External Pseudo-Scalar}

Four-point scattering amplitudes of a pseudo-scalar $\phi$ ($\Gamma=i\gamma_5$) at one-loop can be expressed in terms of the integrals~\eqref{bubble_int}-\eqref{box_int} as 
\begin{align}
\nonumber
M(s,t)&=-2y^4\bigg[ 2\big(I_{\rm bub}(k_1+k_2)+I_{\rm bub}(k_2+k_3)\big)-suI_{\rm box}(-k_1,k_2,k_2+k_3) 
\\
\nonumber
&\qquad\qquad
-s\big(I_{\rm tri}(-k_1,k_2)+I_{\rm tri}(-k_3,k_4)\big)-u\big(I_{\rm tri}(-k_4,k_1)+I_{\rm tri}(-k_2,k_3)\big) \bigg]
\\
&\quad
+\text{2 permutations}\,.
\end{align}
The four- and six-derivative terms of the amplitude are then given by
\begin{align}
M(s,t)=\frac{y^4}{240\pi^2m^4}(s^2+st+t^2)+\frac{y^4}{672\pi^2m^6}st(s+t)\,.
\end{align}
The coefficient of $s^2$ is positive as required by unitarity, whereas the sign of $s^2t$ is positive, hence spinning intermediate states dominate over scalar ones.

\paragraph{External Scalar}
Similarly, one-loop four-point amplitudes of a scalar $\phi$ ($\Gamma=1$) are 
\begin{align}
\nonumber
M(s,t)=-2y^4\bigg[ &2\big(I_{\rm bub}(k_1+k_2)+I_{\rm bub}(k_2+k_3)\big)-(su-32m^4)I_{\rm box}(-k_1,k_2,k_2+k_3) 
\\
\nonumber
&-(s+8m^2)\big(I_{\rm tri}(-k_1,k_2)+I_{\rm tri}(-k_3,k_4)\big)
\\
&-(u+8m^2)\big(I_{\rm tri}(-k_4,k_1)+I_{\rm tri}(-k_2,k_3)\big) \bigg]+\text{2 permutations}\,.
\end{align}
The corresponding four- and six-derivative terms read
\begin{align}
M(s,t)=\frac{11y^4}{720\pi^2m^4}(s^2+st+t^2)-\frac{13y^4}{10080\pi^2m^6}st(s+t)\,.
\end{align}
The coefficient of $s^2$ is positive as required by unitarity, whereas the sign of $s^2t$ is negative, hence scalar intermediate states dominate over spinning ones.

\subsection{On loops for inflation }\label{App_loop}

We have  demonstrated that the sign of the $s^2t$ term generated by loops of heavy fields depends on details of the interactions, essentially because multi-particle states generated by loops may have various spins. While it is interesting that we may use the sign to probe interactions generating loop diagrams, a remark is needed in the context of inflation. As we mentioned, the inflaton enjoys an approximate shift symmetry to respect the slow-roll conditions. Therefore, we may focus on shift-symmetric interactions as far as observable non-Gaussianities are concerned. However, the interactions we discussed in this appendix break the shift symmetry of $\phi$, so that the corresponding non-Gaussianities will be slow-roll suppressed. As far as we know, there are no renormalizable shift-symmetric interactions whose effects appear only at the loop level. For example, the quasi-single field inflation~\cite{Chen:2009zp} accommodates renormalizable interactions of a shift-symmetric scalar (the inflaton) and a massive scalar which source observable non-Gaussianities. However, effects of the massive scalar already appear at the tree-level and thus the loop effects are subdominant. It would be interesting to explore a natural UV model which accommodates dominant loop effects generating observable non-Gaussianities\footnote{Fermion loops generated by higher derivative operators were studied, e.g., in~\cite{Chen:2016uwp,Chen:2018xck}. It would be interesting to construct a UV completion of the effective interactions studied there.}. Even though we leave it for future work, such a direction would be important because the sign of $s^2t$ directly probes spins of heavy particles exchanged at the tree-level if the loop effects are subdominant.

\section{Details on flat space limit}
\label{app:flat}
In this appendix we elaborate on the relation between de Sitter correlators in the flat space limit $k_{1234}\rightarrow 0$ and scattering amplitudes.
As an illustrative example, let us consider four-point interactions of a massless scalar $\phi$ schematically of the form,
\begin{align}
\nonumber
	S_4 &= \frac{a_{pqrs}}{\Lambda^{p+q+r+s}}  \int d^4 x \sqrt{-g} g^{\bullet\bullet} \ldots g^{\bullet\bullet}\, (\nabla_{\mu_1} \ldots \nabla_{\mu_p}\phi)  (\nabla_{\nu_1} \ldots \nabla_{\nu_q}\phi) 
	\\
	&\qquad\qquad\qquad\qquad\qquad\qquad\qquad
	\times (\nabla_{\rho_1} \ldots \nabla_{\rho_r}\phi)  (\nabla_{\sigma_1} \ldots \nabla_{\sigma_s}\phi)~,
\end{align} 
where there are $p+q+r+s$ derivatives in total and their indices have to be contracted by the inverse metric $g^{\bullet\bullet}$ appropriately (we leave $\bullet=\mu_i,\nu_i,\rho_i,\sigma_i$ unspecified). Note that  $p+q+r+s$ should be an even integer. We work in the conformal time coordinates.

The de Sitter four-point function is then given by
\begin{align}\nonumber
& \langle \phi_{\mathbf k_1} (0)  \phi_{\mathbf k_2} (0) \phi_{\mathbf k_3} (0) \phi_{\mathbf k_4} (0)  \rangle  =  (2\pi)^3 \delta (\sum_i \mathbf k_i  )
\\
\nonumber
& \times(- 2) \frac{a_{pqrs}}{\Lambda^{p+q+r+s}} {\rm Im } \bigg[u_{k_1}(0) u_{k_2}(0) u_{k_3}(0) u_{k_4}(0) \int_{-\infty}^0 d\tau (H\tau)^{p+q+r+s-4}  \eta^{\bullet\bullet} \ldots \eta^{\bullet\bullet}
\\
\nonumber
&
\times (\nabla_{\mu_1} \ldots \nabla_{\mu_p} u^*_{k_1}(\tau))  (\nabla_{\nu_1} \ldots \nabla_{\nu_q} u^*_{k_2}(\tau))  (\nabla_{\rho_1} \ldots \nabla_{\rho_r}  u^*_{k_3}(\tau))  (\nabla_{\sigma_1} \ldots \nabla_{\sigma_s}  u^*_{k_4}(\tau))\bigg]
\\
\label{general_int}
&+\text{23 permutations,}
\end{align} 
where $\eta^{\bullet\bullet}$ is the flat space metric and $u_{k_i}(\tau)$ denotes the mode function~\eqref{mode} of $\phi$. Also spatial derivatives should be understood as spatial momenta $\mathbf k_i$.

\medskip
Notice here that the integral~\eqref{general_int} is a linear combination of the integrals,
\begin{align}
\int_{-\infty}^{0} \tau^n e^{i k_{1234} \tau} d \tau = (-1)^nn! \,(ik_{1234})^{-1-n}\,,
\end{align}
with (spatial) momentum-dependent coefficients, so that the highest power in $n$ is dominant in the flat space limit $k_{1234}\to0$. It implies that the contributions from connections are subleading compared with the normal derivative terms in the flat space limit.
We therefore focus on the following part of the integral:
\begin{align}\nonumber
\int_{-\infty}^0 d\tau (H\tau)^{p+q+r+s-4} \eta^{\bullet\bullet} \ldots \eta^{\bullet\bullet} (\partial_{\mu_1} \ldots \partial_{\mu_p} u^*_{k_1}(\tau))  (\partial_{\nu_1} \ldots \partial_{\nu_q}u^*_{k_2}(\tau)) \\ 
\times(\partial_{\rho_1} \ldots \partial_{\rho_r}u^*_{k_3}(\tau))  (\partial_{\sigma_1} \ldots \partial_{\sigma_s}u^*_{k_4}(\tau))~.
\end{align} 
We may further simplify the integral in the flat space limit by using the high energy limit of the mode function,
\begin{align}
	u^*_k (\tau) \rightarrow   -\frac{iH}{\sqrt{2 k}} \tau  e^{ i k \tau}\,.	
\end{align}
Under this approximation, we find
\begin{align}
	\partial_\tau u^*_{k}\rightarrow  -(ik)\times\frac{iH}{\sqrt{2 k}} \tau  e^{i k \tau } ~.
\end{align}
If we think of $k$ as the energy variable and define a four-momentum as
\begin{align}
	k_{\mu} = (-k, \mathbf k)~,
\end{align}
the previous expression becomes
\begin{align}\nonumber
&
\quad
\frac{H^{p+q+r+s}}{4 \sqrt{k_1k_2k_3k_4}}\int_{-\infty}^0 d\tau  \tau^{p+q+r+s} e^{ik_{1234}\tau}
\\
\nonumber
&\quad
\times
i^{p+q+r+s}\eta^{\bullet\bullet} \ldots \eta^{\bullet\bullet} (k_{1\mu_1} \ldots k_{1\mu_p} )( k_{2\nu_1} \ldots k_{2\nu_q})
(k_{3\rho_1} \ldots k_{3\rho_r})  (k_{4\sigma_1} \ldots k_{4\sigma_s})
\\
\nonumber
&=
(p+q+r+s)!\,(ik_{1234})^{-1-(p+q+r+s)}\frac{H^{p+q+r+s}}{4 \sqrt{k_1k_2k_3k_4}}
\\
&\quad\times
i^{p+q+r+s}\eta^{\bullet\bullet} \ldots \eta^{\bullet\bullet} (k_{1\mu_1} \ldots k_{1\mu_p} )( k_{2\nu_1} \ldots k_{2\nu_q})
(k_{3\rho_1} \ldots k_{3\rho_r})  (k_{4\sigma_1} \ldots k_{4\sigma_s})
~,
\end{align} 
where the last line is nothing but what we encounter in the computation of flat space amplitudes. 
In summary, the four-point correlator in the flat space limit yields
\begin{align}\nonumber
	& \langle \phi_{\mathbf k_1} (0)  \phi_{\mathbf k_2} (0) \phi_{\mathbf k_3} (0) \phi_{\mathbf k_4} (0) \rangle \xrightarrow[]{k_{1234}\rightarrow 0}(2\pi)^3 \delta (\sum_i \mathbf k_i  ) \\
	\nonumber
	& \times
	 \frac{a_{pqrs}}{\Lambda^{p+q+r+s}} \frac{H^4}{4 \sqrt{k_1^3 k_2^3 k_3^3 k_4^3}} 2(p+q+r+s)! (k_{1234})^{-1-(p+q+r+s)}\frac{H^{p+q+r+s}}{4 \sqrt{k_1k_2k_3k_4}}
	\\
	&\times
	 \eta^{\bullet\bullet} \ldots \eta^{\bullet\bullet} (k_{1\mu_1} \ldots k_{1\mu_p} )( k_{2\nu_1} \ldots k_{2\nu_q})
	(k_{3\rho_1} \ldots k_{3\rho_r})  (k_{4\sigma_1} \ldots k_{4\sigma_s}) +\text{23 permutations,}
\end{align}
where $\frac{H^4}{4 \sqrt{k_1^3 k_2^3 k_3^3 k_4^3}}$ comes from the contribution of external legs $ u_{k_1}(0) u_{k_2}(0) u_{k_3}(0) u_{k_4}(0)$. The last line gives the flat space amplitude.

\bibliography{sign6d}{}
\bibliographystyle{utphys}

\end{document}